\documentclass[preprint,5p,times,twocolumn,authoryear]{elsarticle}
\usepackage{booktabs}
\usepackage{amsmath,amssymb,amsfonts}
\usepackage{graphicx}
\usepackage{textcomp}
\usepackage{float}
\usepackage{multirow}
\usepackage{pifont}
\usepackage{color}

\journal{Expert Systems with Applications}

\begin{document}

\begin{frontmatter}

\title{DSCA: A Dual-Stream Network with Cross-Attention on Whole-Slide Image Pyramids for Cancer Prognosis \tnoteref{title-lb1}}  

\tnotetext[title-lb1]{This work is supported by the National Natural Science Foundation of China (NSFC) under Grant No. 61972072, partly by Sichuan Science and Technology Program under Grant No. 2021YFS0071 and 2021YFS0236, West China Hospital-UESTC Medical \& industrial integration cross talent training fund ZYGX2022YGRH015, and the West China Hospital 1.3.5 project for disciplines of excellence under Grant No. ZYJC21035.} 

\author[afn1]{Pei Liu}
\ead{yuukilp@163.com}

\author[afn1]{Bo Fu}
\ead{fubo@uestc.edu.cn}

\author[afn2]{Feng Ye}
\ead{fengye@scu.edu.cn}

\author[afn1]{Rui Yang}
\ead{yang19981125@126.com}

\author[afn1]{Luping Ji\corref{cor1}}
\cortext[cor1]{Corresponding author: L. Ji (jiluping@uestc.edu.cn).}
\ead{jiluping@uestc.edu.cn}

\affiliation[afn1]{organization={School of Computer Science and Engineering, University of Electronic Science and Technology of China},
            addressline={Xiyuan Ave}, 
            city={Chengdu},
            postcode={611731}, 
            state={Sichuan},
            country={China}}
\affiliation[afn2]{organization={Institute of Clinical Pathology, West China Hospital, Sichuan University},
            addressline={Guo Xue Xiang}, 
            city={Chengdu},
            postcode={610041}, 
            state={Sichuan},
            country={China}}

\begin{abstract}
The cancer prognosis on gigapixel Whole-Slide Images (WSIs) has always been a challenging task. To further enhance WSI visual representations, existing methods have explored image pyramids, instead of single-resolution images, in WSIs. Despite this, they still face two major problems: high computational cost and the unnoticed semantical gap in multi-resolution feature fusion. 
To tackle these problems, this paper proposes to efficiently exploit WSI pyramids from a new perspective, the dual-stream network with cross-attention (DSCA). Our key idea is to utilize two sub-streams to process the WSI patches with two resolutions, where a square pooling is devised in a high-resolution stream to significantly reduce computational costs, and a cross-attention-based method is proposed to properly handle the fusion of dual-stream features. 
We validate our DSCA on three publicly-available datasets with a total number of 3,101 WSIs from 1,911 patients. 
Our experiments and ablation studies verify that 
(i) the proposed DSCA could outperform existing state-of-the-art methods in cancer prognosis, by an average C-Index improvement of around 4.6\%; 
(ii) our DSCA network is more efficient in computation---it has more learnable parameters (6.31M vs. 860.18K) but less computational costs (2.51G vs. 4.94G), compared to a typical existing multi-resolution network. 
(iii) the key components of DSCA, dual-stream and cross-attention, indeed contribute to our model's performance, gaining an average C-Index rise of around 2.0\% while maintaining a relatively-small computational load. 
Our DSCA could serve as an alternative and effective tool for WSI-based cancer prognosis. Our source code is available at \textit{https://github.com/liupei101/DSCA}.
\end{abstract}


% \begin{highlights}
% \item A novel dual-stream network is proposed to exploit WSI pyramids for cancer prognosis
% \item A square pooling layer is devised to make the network more efficient in computation 
% \item A cross-attention pooling is developed to make dual-stream feature fusion smoother
% \item Our network could perform better than SOTA with an average improvement of 4.6\%
% \item Ablation study verified the effectiveness of our dual-stream and cross-attention
% \end{highlights}

% Keywords
% Each keyword is separated by \sep

\end{frontmatter}

% Main text
\section{Introduction}

\begin{figure}[ht]
\centering
\includegraphics[width=0.48\textwidth]{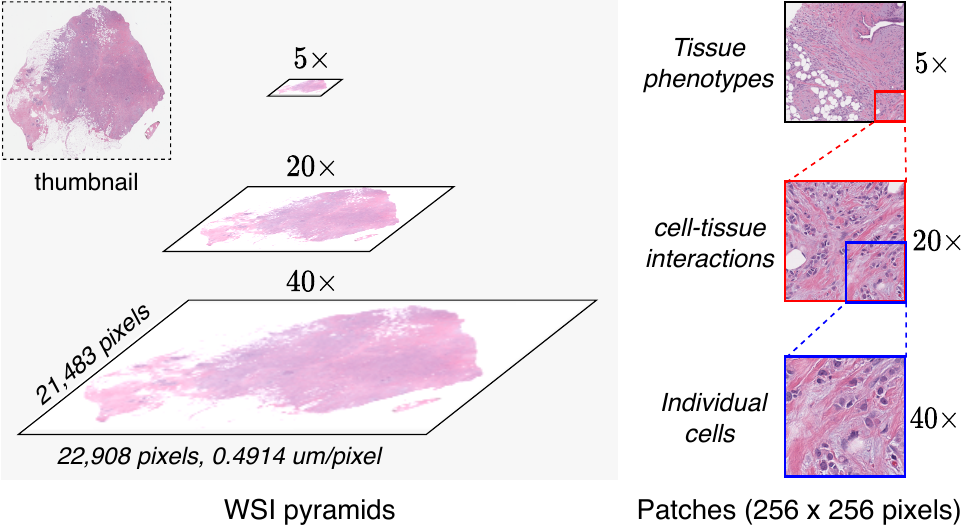}
\caption{Histopathology whole-slide image pyramids and their respective image patches.}
\label{fig0a}
\end{figure}

Cancer prognosis typically answers the question of how the disease will develop for patients. It can be affected by various factors such as primary site, tumor grade, and treatment response \citep{de2009cancer}. Among these factors, some tumor-related ones can be uncovered by the histopathology Whole-Slide Image (WSI) from a high-end microscope \citep{zarella2018apractical,bankhead2017qupath}. Thereby, WSIs are often utilized by doctors to assess patients' diseases and further make clinical decisions. Different from natural images, medical WSIs have extremely-high resolution, usually with gigapixels. And the WSIs at different magnifications present distinct microscopic views \citep{zarella2018apractical, Chen2022scaling}. These views can exhibit rich pathological entities, from tissue phenotypes (5$\times$) to cellular organization (20$\times$), and even to individual cells (40$\times$), as shown in Figure \ref{fig0a}.
\begin{figure*}[tp]
\centering
\includegraphics[width=0.85\textwidth]{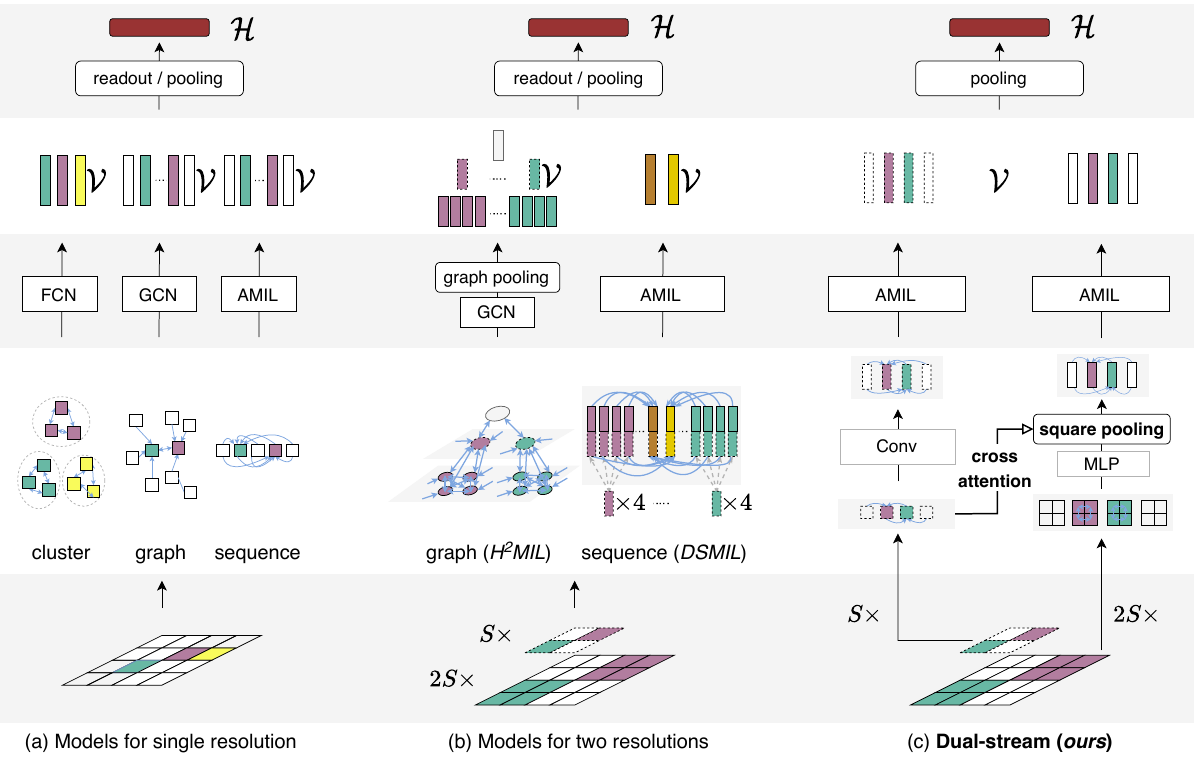}
\caption{Comparison of architectures. Colored patches are taken as examples to illustrate how networks process them. The blue line indicates patch aggregation flow. Dashed ones are low-resolution items. $S$ denotes image resolution. $\mathcal{V}$ is the set of patches after dependency learning. $\mathcal{H}$ denotes WSI-level representation. AMIL means attention-based MIL network and MLP represents multi-layer perceptron. The details of (a) and (b) are available in Section \ref{sec:relatedwork}. The proposed architecture (c) is elaborated in Figure \ref{fig1}.}
\label{fig0b}
\end{figure*}

Due to the gigapixel inherent in WSIs, a computational modeling paradigm \citep{Chen2022scaling} is usually decomposed into (i) slicing images into a myriad of small image patches and (ii) modeling these patches by using bag-level multiple instance learning (MIL) methods. In the early years, most works of modeling WSIs for cancer prognosis focus on single-resolution images \citep{zhu2017wsisa, li2018graph, di2020ranking, yao2020whole, shao2021weakly, chen2021whole, huang2021integration}. From a unified perspective, these works aim at making patch information flow within a specified structure to learn patch dependencies, and then outputting new patch embeddings to derive effective WSI representations, as shown in Figure \ref{fig0b}(a). For example, Li \textit{et al.} originally adopt graph to describe patches \citep{li2018graph}. And they train a graph convolution network (GCN) to learn patch embeddings and WSI representations. Afterward, other patch structures (\textit{e.g.}, cluster and sequence) and networks (\textit{e.g.}, fully-connected networks and attention-based MIL networks) are further tried for WSI-based cancer prognosis \citep{yao2020whole,shao2021weakly,huang2021integration}, as illustrated in Figure \ref{fig0b}(a). However, the single-resolution nature of these schemes greatly hinders models from fully utilizing inherent WSI pyramids (see Figure \ref{fig0a}) for a more accurate cancer prognosis. 

Motivated by pathologist's experience (examining WSIs from the region of interest to tumor localization) \citep{zarella2018apractical}, as well as the idea of hierarchical vision representation \citep{adelson1984pyramid}, WSI-based methods have gradually evolved into multi-resolution from single-resolution over recent years \citep{pati2020hact,li2021dual,hou2022h2,Chen2022scaling}. DSMIL \citep{li2021dual} is a representative multi-resolution scheme. It tries to utilize WSI pyramids by simply concatenating the patch features at different magnifications, as illustrated in Figure \ref{fig0b}(b). By contrast, H$^2$MIL \citep{hou2022h2} adopts a multi-layer graph to model multi-resolution patches. It especially builds cross-layer connections (or edges) for cross-resolution patches to integrate and exploit hierarchical features, as shown in Figure \ref{fig0b}(b). Although these multi-resolution schemes have demonstrated better results than single-resolution ones, the way of \textit{explicitly} using high-resolution patches throughout patch embedding learning is not efficient enough. Because there is usually an enormous amount of patches in high-resolution WSIs, whose number may be up to 10,000 at 20$\times$ magnification. In addition, from the perspective of multi-scale feature fusion \citep{lin2017feature,tan2020efficientdet,zhang2020feature}, existing schemes have not paid enough attention to the intrinsic semantical gap in different resolution patches, but are restricted to simple concatenating (DSMIL) or direct message passing (H$^2$MIL) for multi-resolution feature fusion. 

To tackle the problems analyzed above, we propose an alternative scheme based on a different idea, a dual-stream network with cross-attention (DSCA), as shown in Figure \ref{fig0b}(c). This scheme is inspired by modern feature pyramid networks \citep{lin2017feature,tan2020efficientdet,zhang2020feature,wang2021pyramid}. Briefly, our multi-resolution scheme consists of two vital components: dual-stream and cross-attention. The former is utilized to process low- and high-resolution patches respectively, so as to efficiently learn hierarchical WSI representation; the latter handles the potential semantical gap in dual-resolution features. Specifically, in the high-resolution stream, we devise a square pooling layer to largely decrease the number of high-resolution patches, exploiting fine-grained features efficiently. Furthermore, this square pooling is implemented by a cross-attention-based method that pools high-resolution patches under the guidance of a global low-resolution one. Notably, our DSCA could be easily extended to multi-stream to model multi-resolution WSIs when its computational cost is feasible. 

We summarize our main contributions as follows:

(1) A dual-stream network with cross-attention (DSCA) is proposed to learn hierarchical representations from WSI pyramids for cancer prognosis. It has three highlights: 
\begin{itemize}
    \item We propose a novel scheme, a dual-stream network with cross-attention, to fully utilize image pyramids for enhancing the visual representation of WSIs. This scheme's key idea is quite distinct from other WSI-based multi-resolution ones. 
    \item A square pooling layer is devised to significantly decrease the number of high-resolution patches, thus largely reducing the computational cost in network training. It serves as a critical component for computation efficiency. 
    \item A cross-attention-based pooling method is proposed to pool high-resolution patches under the guidance of a global low-resolution one. It could effectively handle the potential semantical gap between high- and low-resolution features, making the fusion of dual-stream features smoother. It is a specific implementation of square pooling leveraged to enhance hierarchical representation for better prediction performance.
\end{itemize} 

(2) We carry out extensive experiments on three publicly-available WSI datasets and empirically prove that the proposed DSCA could often outperform existing state-of-the-art methods in cancer prognosis owing to its dual-stream architecture with cross-attention.

\begin{table*}[tp]
\small
\centering
\caption{Works related to WSI classification or cancer prognosis.}
\label{tb1}
\begin{tabular}{cccccc}
\toprule
Structure & Method & Task & Pyramid &  Network & Highlights \\ \midrule
\multirow{3}{*}{cluster}        & WSISA \citep{zhu2017wsisa}           & prognosis &            & FCN           & the first deep learning model       \\
                                & DeepAttnMISL \citep{yao2020whole}    & prognosis &            & FCN           & cluster attention \\
                                & BDOCOX \citep{shao2021weakly}        & prognosis &            & FCN           & pseudo-bag; ranking loss   \\ \cmidrule{1-1}
\multirow{4}{*}{graph}          & DeepGraphSurv \citep{li2018graph}    & prognosis &            & SpectralGCN   & the first graph-based model   \\
                                & RankSurv \citep{di2020ranking}       & prognosis &            & HyperGCN      & hyper-graph network    \\
                                & PatchGCN \citep{chen2021whole}       & prognosis &            & GCN           & regional graph                \\ 
                                & H$^2$MIL \citep{hou2022h2}           & classification & \ding{51}  & GCN           & heterogeneous graph               \\\cmidrule{1-1}
\multirow{5}{*}{sequence}          & SeTranSurv \citep{huang2021integration} & prognosis &            & Transformer   & Transformer MIL   \\
                                & DT-MIL \citep{li2021dt}          & classification &       & Transformer   & deformable Transformer MIL          \\
                                & TransMIL \citep{shao2021transmil}        & classification &       & Transformer   & CNN positional encoding             \\
                                & DSMIL \citep{li2021dual}           & classification &  \ding{51}     & Attention-MIL & the first patch pyramid model             \\
                                & DTFD-MIL \citep{zhang2022dtfd}        & classification &       & Attention-MIL & CAM inspired MIL aggregation        \\ 
                                & HIPT \citep{Chen2022scaling}           & both &  \ding{51}     & Transformer & self-supervised pre-training             \\ %\midrule
% sequence & \textbf{DSCA} (ours) & prognosis & \ding{51} & Transformer & dual-stream; cross-attention \\ 
\bottomrule
\end{tabular}
\end{table*}

\section{Related work}
\label{sec:relatedwork}

This section will introduce (i) the image and feature pyramids in computer vision, (ii) traditional multiple instance learning (MIL) framework, and (iii) MIL-based WSI representation learning. To better compare the existing works related to WSI representation learning, we summarize representative ones and exhibit them in Table \ref{tb1}. 

\subsection{Image and feature pyramid}
In computer vision, feature pyramids also refer to multi-scale or hierarchical features. It can be extracted from images to enhance visual representation. Traditional computer vision methods usually get feature pyramids from image pyramids by scaling original images \citep{adelson1984pyramid}. By contrast, modern deep learning methods typically achieve this purpose by only using ConvNets \citep{simonyan2014very,he2016deep}, not directly relying on image pyramids. Because ConvNets can produce feature maps with multi-scale receptive fields, showing a pyramidal shape. And these feature maps are inherent with hierarchical and semantically-distinct information \citep{zeiler2014visualizing}. 

In order to effectively exploit the feature pyramids derived from the nature of deep neural networks, a typical solution \citep{lin2017feature,tan2020efficientdet,zhang2020feature} is to make multi-scale feature maps in the same scale via downsampling or upsampling operators, and then to add these feature maps for a multi-scale feature fusion. However, as the approaches that use feature pyramids similarly, existing multi-resolution schemes have not paid enough attention on how to better fuse multi-resolution patch features in WSI representation learning, but are restricted to a simple feature concatenating \citep{li2021dual} or a direct message (feature) passing between graph nodes \citep{hou2022h2}. %By contrast, inspired by the typical solution in modern computer vision, our dual-stream scheme proposes to achieve the effective fusion of multi-resolution patches by using square pooling and cross-attention for gigapixel WSIs. 

\subsection{Multiple instance learning}

Multiple Instance Learning (MIL) is a weakly-supervised learning framework \citep{dietterich1997solving,carbonneau2018multiple}. In MIL settings, the dataset can be written by  
\begin{equation}
\begin{gathered}
T=\bigl\{(\mathcal{X}_i,y_i)\bigr\}_{i=1}^{i=N},\ \mathcal{X}_i=\bigl\{I_i^{(j)}\bigr\}_{j=1}^{j=n_i},
\end{gathered}
\label{eq0}
\end{equation}
where $\mathcal{X}_i$ denotes a bag sample, $y_i$ is the label of $\mathcal{X}_i$, and $I_i^{(j)}$ is the $j$-th instance of $\mathcal{X}_i$. A bag sample contains multiple instances and only bag labels are available. Traditional MIL approaches can be divided into bag-level MIL and instance-level MIL \citep{wang2018revisiting}. And the former aims to learn bag-level representation. 

In WSI modeling, WSIs and patches can be viewed as bags and instances, respectively. Thus, most WSI-based approaches are also cast as MIL. Moreover, these approaches belong to bag-level MIL, focusing on learning WSI-level representation. We will review them in the following part. 

\subsection{WSI-level representation learning}
\label{subsubsec:amilmethod}

\subsubsection{Single-resolution}
The schemes in this class usually use the WSIs at 20$\times$ magnification, since this resolution is enough to capture cell-tissue interactions and its computational cost is reasonable. To learn WSI-level representation, these schemes often learn patch dependencies at first and then aggregate all patch embeddings into WSI-level features, as shown in Figure \ref{fig0b}(a). In patch dependency learning, patches are often organized as clusters, graphs, or sequences (referring to those patches without any specified structures). (i) Cluster-based schemes \citep{zhu2017wsisa,yao2020whole,shao2021weakly} often adopt patch similarity for clustering and then learn patch dependencies in each cluster using fully-connected networks. (ii) Graph-based schemes \citep{li2018graph,di2020ranking,chen2021whole} turn to adopt graph to construct patch dependences. They pass or aggregate patch-level features along graph edges using GCNs. (iii) Sequence-based schemes \citep{huang2021integration,li2021dt,shao2021transmil} don't limit any structure for patches, \textit{i.e.}, their models are structure-free. They learn all possible dependencies via attention-based networks, \textit{e.g.}, Transformer \citep{vaswani2017attn}, Attention-MIL \citep{ilse2018attn}, or Vision Transformer (ViT) \citep{dosovitskiy2020vit}. WSI-level representation is usually obtained from patch-level embeddings by a global pooling operator at the end. However, the single-resolution nature of these various schemes could hinder models from fully utilizing WSI pyramids (see Figure \ref{fig0a}). 

\subsubsection{Multi-resolution}
Both DSMIL and H$^2$MIL belong to this class, as shown in Figure \ref{fig0b}(b). In sequence-based DSMIL, multi-resolution features are simply concatenated for patch dependency learning. And only those selected key patches could have connections with other patches. In graph-based H$^2$MIL \citep{hou2022h2}, the edges between low- and high-resolution patches are built additionally, compared to those graph-based single-resolution models. However, these cross-resolution edges only can be used to transfer patch features directly, and the intrinsic semantical gap between different resolution patches remains when calculating WSI-level features by global patch pooling. Most recently, the HIPT \citep{Chen2022scaling}, built upon ViT, also utilizes WSI pyramids and has achieved promising results on WSI classification and prognosis, but it focuses on studying a hierarchical self-supervised pre-training strategy for WSIs. %which is different from the topic of this paper. 

It is worth noting that, among the approaches exhibited in Table \ref{tb1}, DSMIL and DTFD-MIL \textit{only can be used for classification}, because DSMIL selects key patches using class scores and DTFD-MIL relies on class-specific CAMs \citep{zhou2016learning}. In addition, as the two most popular tasks on WSIs---classification and prognosis, there are many differences between them. For example, identifying a specific tumor stage or subtype is usually the end of most classification-oriented models \citep{moitra2020classification,houssein2021deep,zeiser2021deepbatch,pati2022hier,ding2022fractal}. Although tumor stage and subtype are well-known prognostic factors, more evidence in WSIs, such as the certain traits of cancer cells and the tumor surrounding micro-environmental cues \citep{roy2011role,yu2016pred}, is often required to gather by cancer prognosis models \citep{shao2019inter,skrede2020deep,liu2022eoca}. This means that actual prognostic patterns may be more complex in patients’ WSIs. And these factors can vary across cancer types and individuals due to tumor heterogeneity and individual diversity \citep{fu2019predict,alizadeh2015toward}, thus leaving the cancer prognosis on WISs as a more difficult task. We will discuss more through empirical experiments. 

\section{Methodology}

\begin{figure*}[tp]
\centering
\includegraphics[width=0.88\textwidth]{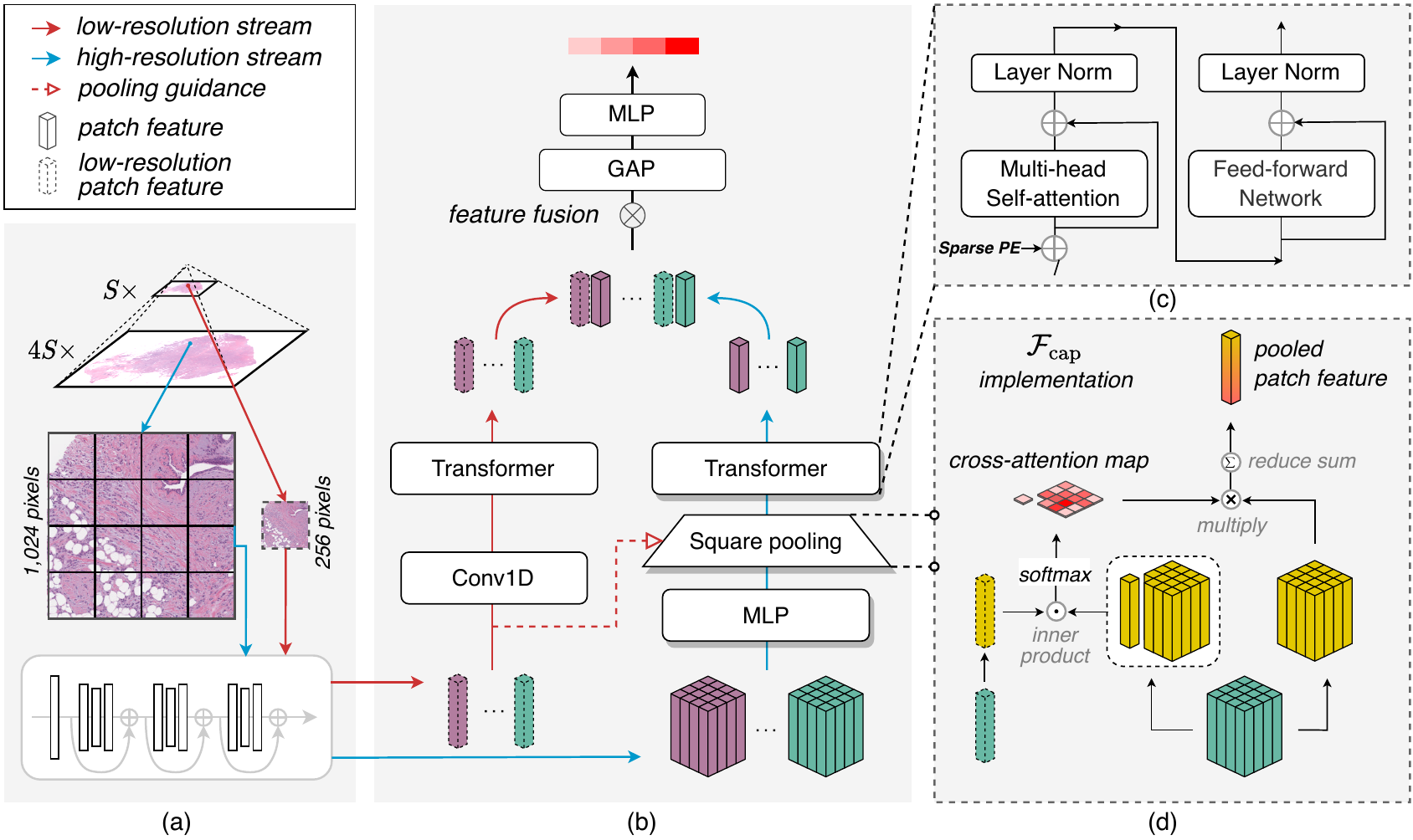}
\caption{Overview of our DSCA scheme (best viewed in color). We illustrate our scheme using $\lambda=4$ and $a=256$. (a) Spatially-aligned dual-resolution patching: $S\times$ and $4S\times$ images are sliced into non-overlapping patches and their features are extracted by a pre-trained model. (b) DSCA architecture. (c) Transformer encoder in DSCA. (d) Cross-attention pooling $\mathcal{F}_{\mathrm{cap}}$: it's an implementation of square pooling that pools high-resolution tokens under the guidance of low-resolution ones.}
\label{fig1}
\end{figure*}

Our DSCA architecture (see Figure \ref{fig1}) has four key components: 1) token embedding layer, 2) dependency learning layer, 3) dual-stream fusion, and 4) prediction layer, from input to output. The first two components are dual-stream versions, as shown in Figure \ref{fig1}(b). The network architecture of two sub-streams roughly follows the philosophy of ViT \citep{dosovitskiy2020vit}. 

\subsection{Problem formulation}

\subsubsection{Dataset}
We denote our dataset by 
\begin{equation}
\begin{gathered}
T=\bigl\{(\mathcal{P}_i,t_i,c_i)\bigr\}_{i=1}^{i=N},\ \mathcal{P}_i=\bigl\{p_i^{(j)}\bigr\}_{j=1}^{j=m_i},
\end{gathered}
\label{eq1}
\end{equation}
where $\mathcal{P}_i, t_i,\text{ and }c_i$ are the WSI material, follow-up time, and censorship status of the $i$-th patient, respectively. $p_i^{(j)}$ denotes the $j$-th patch data in $\mathcal{P}_i$. And $m_i$ is the total number of patches in $\mathcal{P}_i$, which could be different across patients. Throughout the paper, the word \textit{patch} means the \textit{image patch} sliced from WSIs, and the word \textit{token} means the \textit{feature vector} corresponding to that patch. 

\subsubsection{Dual-resolution}
The dual-resolution patches from WSI pyramids are used for our DSCA network. Without loss of generality, we denote the magnification of patches with a lower resolution by $S\times$, and that with higher resolution by $\lambda S\times$, where the typical values of $\lambda$ could be 2, 4, or 8. The subscript $l$ and $h$ indicate \textit{low}-resolution and \textit{high}-resolution, respectively. 

\subsubsection{WSI-level cancer prognosis}
Only with WSI-level labels, DSCA needs to learn WSI-level representations from multiple dual-resolution patches for cancer prognosis prediction in a weakly-supervised manner. Similar to bag-level MIL, WSI-level representation learning also has two essential steps: patch embedding learning and global patch pooling (or aggregation). 

\subsection{Spatially-aligned dual-resolution patching}
First of all, we derive patches from the WSIs at two magnifications. The illustration of WSI pre-processing can be found in Figure \ref{fig1}(a). Specifically, for one WSI, we slice it at a low resolution ($S\times$) and obtain the image patches with $a\times a$ pixels ($a$ is 256 usually). Then we slice the same WSI at a high resolution ($\lambda S\times$), generating patches with $a\times a$ pixels similarly. Moreover, as shown in Figure \ref{fig1}(a), a $S\times$ patch is exactly spatially-aligned to the $\lambda^2$ patches at $\lambda S\times$ (\textit{i.e.}, a square grid of $\lambda\times\lambda$ cells). Lastly, we use a truncated ResNet-50 model \citep{he2016deep} pre-trained on ImageNet \citep{deng2009imagenet} to extract instance features from each patch, following \citep{lu2021data}. 

Given one patient with multiple WSIs, we denote its processed patch tokens at $S\times$ and $\lambda S\times$ by  
$$
\mathrm{\mathbf{X}}_{l}\in \mathbb{R}^{m\times d} \text{ and } \mathrm{\mathbf{X}}_{h}\in \mathbb{R}^{\lambda^2m\times d},
$$
respectively, where $m$ is the number of $S\times$ patches and $d$ is the dimensionality of token. Comparing to $\mathrm{\mathbf{X}}_{h}$, $\mathrm{\mathbf{X}}_{l}$ contains coarser features and has a larger receptive field. Conversely, $\mathrm{\mathbf{X}}_{h}$ can provide more local and fine-grained information. This complementary nature helps to enhance the visual representation of gigapixel WSIs. 

\subsection{Dual-stream token embedding}

\subsubsection{Low-resolution stream}
For this stream, we adopt a one-dimensional convolution layer as its token embedding layer for simplicity. $\mathrm{\mathbf{X}}_{l}$ is padded with zeros to keep the resolution. We write the output of this convolution layer by  
\begin{equation}
    \mathrm{\mathbf{E}}_{l}=\mathrm{Conv1D}\bigl(\mathrm{\mathbf{X}}_{l};k,s\bigr),
    \label{eq2}
\end{equation}
where $\mathrm{\mathbf{E}}_{l}\in \mathbb{R}^{m\times d_e}$ and $d_e$ ($<d$) is the dimensionality of token embedding. $k\text{ and }s$ denote kernel size and stride length, respectively. $\mathrm{Conv1D}$ also plays a role of dimension reduction in token embedding. Moreover, 1D-convolution could integrate the individual patches within overlapping convolution windows, hence outputting overlapping token embeddings. This way could help to obtain more local continuity in tokens, as demonstrated in \citep{wang2021pyramid}. 

\subsubsection{High-resolution stream}
\label{subsubsec:hstream}
In the high-resolution stream used for $\lambda S\times$ patches, we take all the tokens in a $\lambda\times\lambda$ square grid as a unit, and employ a token embedding operator for all these units. Specifically, we firstly reshape $\mathrm{\mathbf{X}}_{h}$ into square sequences, which is implemented by the function, $resize: \mathbb{R}^{\lambda^2m\times d}\to \mathbb{R}^{m\times \lambda\times \lambda\times d}$. Then, the reshaped input goes through a token embedding layer and an activation layer in turn, written as 
\begin{equation}
\begin{gathered}
    \mathrm{\mathbf{O}}_{h} = \sigma\bigl(\rho\bigl(resize(\mathrm{\mathbf{X}}_{h})\bigr)\bigr), \\
    \rho: \mathbb{R}^{m\times \lambda\times \lambda\times d}\to \mathbb{R}^{m\times \lambda\times \lambda\times d_e},
\end{gathered}
\label{eq3}
\end{equation}
where $\sigma$ is an activation function and $\mathrm{\mathbf{O}}_{h}\in \mathbb{R}^{m\times \lambda\times \lambda\times d_e}$. We utilize a simple implementation, MLP (multi-layer perceptron), for $\rho$. MLP calculates each token embedding individually, thereby generating compact token embeddings without keeping local continuity. It is worth noting that, here we do not adopt a convolution-based implementation to produce overlapping embeddings, like that in the low-resolution stream. It is because, in the high-resolution stream, overlapping embedding may have greater potential to incur more information redundancy, as gigapixel WSIs have obviously-lower information density than natural images. We will further discuss this in experiments. 

\subsection{Square pooling with cross-attention}
After high-resolution token embedding, we further apply a square pooling layer implemented by cross-attention to high-resolution tokens ($\mathrm{\mathbf{O}}_{h}$), for the two purposes: (i) significantly decreasing the number of high-resolution tokens and enabling the closely-followed Transformer to learn token dependency efficiently; (ii) handling the potential semantical gap between low- and high-resolution tokens via a cross-attention mechanism, so as to make the fusion of two resolution tokens smoother. 

Specifically, our token pooling operator is similarly applied to square units, defined by $\mathcal{F}:\mathbb{R}^{m\times \lambda\times \lambda\times d_e}\to \mathbb{R}^{m\times d_e}$. As illustrated in Figure \ref{fig1}(d), we implement a cross-attention-based method for $\mathcal{F}$, written as $\mathcal{F}_{\mathrm{cap}}$. In short, $\mathcal{F}_{\mathrm{cap}}$ uses the attention scores mapped from dual-stream tokens to perform a weighted pooling for high-resolution tokens. Namely, our square pooling with cross-attention is implemented by 
\begin{equation}
\begin{gathered}
    \mathrm{\mathbf{Q}}_{l} = \mathrm{\mathbf{X}}_{l}\mathbf{W}_l,\\
    \mathrm{\mathbf{E}}_{h} = \mathcal{F}_{\mathrm{cap}}(\mathrm{\mathbf{O}}_{h};\mathrm{\mathbf{Q}}_{l}),
\end{gathered}
\label{eq4}
\end{equation}
where $\mathrm{\mathbf{E}}_{h}\in \mathbb{R}^{m\times d_e}$ and $\mathbf{W}_l\in \mathbb{R}^{d\times d_e}$. $\mathbf{W}_l$ is a projection matrix to make the token dimension of $\mathrm{\mathbf{Q}}_{l}$ same to $\mathrm{\mathbf{O}}_{h}$. Moreover, taking a low-resolution token (denoted as $\mathbf{z}_{l}\in \mathbb{R}^{1\times d_e}$) and its respective high-resolution tokens (denoted as $\mathbf{z}_{h}\in \mathbb{R}^{\lambda^2\times d_e}$) as examples, $\mathcal{F}_{\mathrm{cap}}$ can be written by  
\begin{equation}
    \mathcal{F}_{\mathrm{cap}}(\mathbf{z}_{h};\mathbf{z}_{l})=\mathrm{softmax}\left(\frac{\mathbf{z}_{l}\mathbf{W}_q(\mathbf{z}_{h}\mathbf{W}_k)^\mathrm{T}}{\sqrt{d_e}}\right) \mathbf{z}_{h}\mathbf{W}_v,
\label{eq5}
\end{equation}
where $\mathbf{W}_q,\mathbf{W}_k,\mathbf{W}_v\in \mathbb{R}^{d_e\times d_e}$ are the projection matrices for query, key, and value, respectively. 

Our cross-attention scheme does spatial attention pooling for large region ($\lambda a\times\lambda a$ pixels). It enables high-resolution tokens with local and fine-grained features to pool under the guidance of a token with a global view. Therefore, the result of cross-attention pooling is expected to selectively preserve certain fine-grained information based on the spatial response of high-resolution patches to their global view. In addition, with the square pooling, the number of high-resolution tokens can be decreased by a factor of $\frac{1}{\lambda^2}$, becoming identical to the number of low-resolution ones. 

\subsection{Transformer-based token dependency learning}

\subsubsection{Sparse positional embedding}
\label{subsubsec:pe}
We calculate the positional embedding (PE) of tokens similar to Transformer \citep{vaswani2017attn}, rather than ViT \citep{dosovitskiy2020vit}, because, unlike general cases, the size of tokens (\textit{i.e.}, the value of $m$) varies across WSIs. In addition, there is a problem of spatially-sparse distribution in tokens, as we filter the patches without any tissue region in WSI patching. 

To solve this, we put all WSIs from one patient in a row and then map the coordinates of all available patches to a discrete domain without breaking their relative locations. In this way, we can encode the position information of tokens while preserving intrinsic spatially-sparse property, even for multiple WSIs. For simplicity, we write the result of sparse PE as $$\mathrm{\mathbf{PE}}\in \mathbb{R}^{m\times d_e}.$$ 

\subsubsection{Dependency learning}
As shown in Figure \ref{fig1}(c), we employ a Transformer encoder at the end of either stream to learn potential token dependencies, as Transformer has shown great success in capturing token dependencies \citep{vaswani2017attn}. This encoder is a vanilla Transformer encoder that consists of a multi-head self-attention layer and a feed-forward network. We denote these two parts by $\mathrm{MSA}: \mathbb{R}^{m\times d_e}\to \mathbb{R}^{m\times d_e}$ and $\mathrm{FFN}: \mathbb{R}^{m\times d_e}\to \mathbb{R}^{m\times d_e}$, respectively. The Transformer encoder, $\mathcal{T}: \mathbb{R}^{m\times d_e}\to \mathbb{R}^{m\times d_e}$, can be expressed by 
\begin{equation}
\begin{gathered}
    \mathcal{T} = \zeta \circ \varphi, \\
    \zeta(x) = \mathrm{LN}\bigl(\mathrm{MSA}(x) + x\bigr),\\ 
    \varphi(x) = \mathrm{LN}\bigl(\mathrm{FFN}(x) + x\bigr),
\end{gathered}
\label{eq6}
\end{equation}
where $\mathrm{LN}$ is a layer normalization function \citep{ba2016layer}. Based on Equation \ref{eq6}, we can get the outputs as follows: $$\mathrm{\mathbf{V}}_{l}=\mathcal{T}_{l}(\mathrm{\mathbf{E}}_{l} + \mathrm{\mathbf{PE}}) \text{ and } \mathrm{\mathbf{V}}_{h}=\mathcal{T}_{h}(\mathrm{\mathbf{E}}_{h} + \mathrm{\mathbf{PE}}).$$ 

\subsection{Dual-stream feature fusion}
\label{subsubsec:ff}
Stream feature fusion (FF) is implemented by the following function: 
\begin{equation}
    \mathrm{\mathbf{F}} = \mathrm{FF}(\{\mathrm{\mathbf{V}}_{l}, \mathrm{\mathbf{V}}_{h}\}).
\label{eq7}
\end{equation}
We use a simple implementation for $\mathrm{FF}$, feature concatenating, defined by $$\mathrm{cat}: \mathbb{R}^{2m\times d_e}\to \mathbb{R}^{m\times 2d_e}.$$ In addition, we consider an adding version for $\mathrm{FF}$. Their results are shown in experiments. 

\subsection{Prognosis prediction}
\subsubsection{Output layer}
\label{subsubsec:outlayer}
We utilize a global attention pooling (GAP) \citep{ilse2018attn} to obtain WSI-level features. Specifically, all available tokens are weighted by their respective attention scores, and then they are aggregated into a global vector. This global vector is finally employed to infer prognosis prediction via a multi-layer perceptron (MLP). Thus, our output layer implements 
\begin{equation}
\begin{gathered}
    \mathrm{\mathbf{H}} = \mathrm{GAP}(\mathrm{\mathbf{F}}), \\
    \mathrm{\mathbf{O}} = \mathrm{sigmoid}\bigl(\mathrm{MLP}(\mathrm{\mathbf{H}})\bigr),
\end{gathered}
\label{eq8}
\end{equation}
where $\mathrm{GAP}: \mathbb{R}^{m\times d_o}\to \mathbb{R}^{d_o}$, $\mathrm{MLP}: \mathbb{R}^{d_o}\to \mathbb{R}^{n_{t}}$, and $\mathrm{\mathbf{O}} \in \mathbb{R}^{n_{t}}$ is the prediction of individual hazard function. $d_o$ is the dimensionality of WSI-level features, which equals $2d_e$. $n_{t}$ denotes the number of time points considered in survival prediction. 

\subsubsection{Loss function}
The loss function for optimizing our DSCA network is a negative log-likelihood function \citep{zadeh2020bias} used for training discrete deep survival models:  
\begin{equation}
\begin{gathered}
    L = L_{\text{uncensored}} + (1-\alpha)L_{\text{censored}},\\
    L_{\text{uncensored}} = -\sum\limits_{(\mathcal{P}_i,t_i,c_i)\in T}\Bigl\{(1-c_i) \cdot \mathrm{log}\bigl(S(t_i-1)\cdot h(t_i)\bigr)\Bigr\}, \\
    L_{\text{censored}} = -\sum\limits_{(\mathcal{P}_i,t_i,c_i)\in T}\Bigl\{c_i\cdot \mathrm{log}\bigl(S(t_i)\bigr)\Bigr\},
\end{gathered}
\label{eq9}
\end{equation}
where $\alpha\in \mathbb{R}_{0}^{+}$, $h(t_i)$ is the $t_i$-th element of the network prediction $\mathrm{\mathbf{O}}$, and $S(t_i)=\prod_{s=1}^{t_i}\bigl(1-h(s)\bigr)$. In survival analysis, $h(t)$ is a hazard function, and $S(t)$ is a survival function indicating the probability of surviving time $t$. This survival function does not rely on any assumption about the form of hazard function. By contrast, the popular loss functions \citep{cox1975partial, liu2021opt}, based on Cox model, have a strong assumption---proportional hazard. We thereby adopt Equation \ref{eq9} to optimize DSCA, as suggested in \citep{zadeh2020bias}. 

\section{Experiments and results}
\subsection{Datasets}
We validate our DSCA on the three large publicly-available datasets: National Lung Screening Trial (NLST) \citep{nlst2011}, BReast CAncer (BRCA), and Low-Grade Glioma (LGG). The last two datasets are from The Cancer Genome Atlas \citep{Kandoth2013MutationalLA}. As shown in Table \ref{tb2}, we obtain a total of 3,101 WSIs from 1,911 patients, after excluding unavailable WSIs and the patients with unknown follow-up status. Note that there is not any other form of data curation except that we mentioned. 

\begin{table}[tp]
\small
\centering
\caption{Statistic of datasets. The number of patches is measured after slicing WSIs at 20$\times$ magnification.}
\label{tb2}
\begin{tabular}{cccc}
\toprule
\multirow{2}{*}{Name}       & \multirow{2}{*}{NLST} & \multicolumn{2}{c}{TCGA}     \\ \cmidrule(lr){3-4} 
                            &                       & BRCA          & LGG          \\ \midrule
primary site & lung & breast & brain \\
storage size & 0.76 TB     & 0.98 TB       & 0.88 TB      \\
death ratio  & 35.9\%      & 13.5\%        & 23.4\%       \\
\# patients  & 447         & 978           & 486          \\
\# WSIs     & 1,222       & 1,043         & 836          \\ 
\# patches (20$\times$)  & 3,955,344   & 3,228,480     & 2,637,456    \\
$\frac{\text{\# patches}}{\text{\# WSIs}}$   (20$\times$) & 3,236.8   & 3,095.4     & 3,154.9    \\ \bottomrule
\end{tabular}
\end{table}

From Table \ref{tb2}, we can see that the overall death ratio of breast cancer is apparently smaller than that of the other two cancer diseases. The prognosis event of interest that we predict is the overall survival (OS), a typical endpoint in survival analysis. We can notice that there are usually more than 3,000 patches per WSI at 20$\times$ on average. 

\begin{table*}[tp]
\small
\centering
\caption{Prognosis prediction performance measured by 5-fold cross-validation. The asterisk (*) indicates that the method was originally proposed for WSI classification. Baseline comparisons are elaborated in Section \ref{subsubsec:baseline}. The \# MACs of model are measured using a bag with 210 patches at low resolution and 3,360 patches at high resolution. We use the same patch feature extractor and loss function in all methods for fair comparisons.}
\label{tb3}
\begin{tabular}{cccccc}
\toprule
\multirow{2}{*}{Method}    & \multicolumn{3}{c}{Mean C-Index (Standard Deviation)} & \multirow{2}{*}{Model Size} & \multirow{2}{*}{\# MACs} \\ \cmidrule(lr){2-4}
 & NLST & BRCA & LGG & & \\ \midrule
Deep Sets \citep{zaheer2017deep}   & 0.47271 (0.08325)                                                       & 0.47106 (0.03884) & 0.60080 (0.07250) & 657.16K & 1.76G \\
Attention-MIL \citep{ilse2018attn}   & 0.57157 (0.09022)                                                     & 0.54118 (0.06273) & 0.60820 (0.04624) & 920.07K & 2.64G      \\
DeepAttnMISL \citep{yao2020whole}   & 0.55062 (0.06372)                                                    & 0.53234 (0.03403) & 0.61963 (0.05941) & 1.18M & 1.77G                  \\
DeepGraphConv \citep{li2018graph}   & 0.57931 (0.06061)                                                     & 0.54254 (0.07946) & 0.61954 (0.07918) & 790.02K & 2.42G             \\
Patch-GCN \citep{chen2021whole}      & 0.58588 (0.08245)                                                     & 0.49325 (0.02457) & 0.67261 (0.01605)  & 1.38M & 3.75G            \\
SeTranSurv \citep{huang2021integration}    & 0.54203 (0.10244)                                               & 0.55339 (0.09756) & 0.59912 (0.03948) & 1.58M & 1.22G            \\ 
TransMIL\textsuperscript{\rm *} \citep{shao2021transmil}    & 0.59535 (0.03053)                     & 0.53175 (0.06944) & 0.69049 (0.05121) & 2.67M & 9.55G                          \\ 
H$^2$MIL\textsuperscript{\rm *} \citep{hou2022h2}   & 0.60129 (0.03948)                     & 0.53125 (0.05582) & 0.56867 (0.11425)  &  860.18K & 4.94G               \\ 
\midrule
\textbf{DSCA} (ours) & \textbf{0.66803} (0.05817) & \textbf{0.61224} (0.05350) & \textbf{0.70181} (0.04321) & 6.31M  & 2.51G \\ 
\bottomrule
\end{tabular}
\end{table*}

\subsection{Experimental setup}
\subsubsection{Evaluation metrics}
We adopt Concordance Index (C-Index) \citep{heagerty2005survival} to measure model performance. C-Index is a widely-used evaluation metric in survival analysis, which mainly assesses the model ability of survival risk discrimination. It is written as $$\text{C-Index} = \frac{1}{M}\sum_{i:c_i=0}\sum_{j:t_i<t_j}\mathcal{I}\bigl(\hat{y}_i > \hat{y}_j\bigr)\in [0,1],$$ where $\mathcal{I}(\cdot)$ is an indicator function and $M$ is the number of all comparable pairs. We compute $\hat{y}$ by a negative average over the survival probabilities at all time points, \textit{i.e.}, $\hat{y}=\frac{1}{n_t}\sum_{k=1}^{n_t}S(k)$. The C-Index would be larger if a model could predict higher risks for the patients who die earlier. 

\subsubsection{Model evaluation}
We use 5-fold cross-validation to evaluate all the models in our experiments. In each fold training, we retain the 20\% training set as a validation set for early stopping and learning rate adjusting. Data splitting is ensured to conduct at patient-level. To assess the computation efficiency of models in space and time complexity \citep{tan2020efficientdet}, we report the size of trainable parameters (Model Size) and the theoretical amount of Multiply–Accumulate operations (\# MACs) in models, respectively, by using a package \textit{ptflops}\footnote{https://pypi.org/project/ptflops}. 

\subsubsection{Dual-resolution setup}
We adopt 20$\times$ for high resolution in our experiments, because this resolution is a common choice in previous literature, \textit{e.g.}, \cite{li2018graph,di2020ranking,chen2021whole,shao2021weakly,huang2021integration,li2021dt,shao2021transmil,li2021dual,zhang2022dtfd,Chen2022scaling}. Moreover, this resolution could capture cell-tissue interactions for downstream prediction, and its computational cost is reasonable. In addition, we select $5\times$ for low resolution, \textit{i.e.}, $\lambda=4$, as this resolution could present tissue phenotype, contributing to prognosis prediction \citep{Chen2022scaling}, as shown in Figure \ref{fig0a}.

\subsubsection{Implementation details}
For WSI patching, we set the size of patch to $256\times 256$ pixels, \textit{i.e.}, $a=256$, following \cite{lu2021data}. And $d=1024$, output by a truncated ResNet-50 model. In Conv1D, we empirically set $k=5$ and $s=3$. For each dataset we use the same settings: 150 epochs, a learning rate of $8\times 10^{-5}$, a gradient accumulation step of 16, an optimizer of Adam with a weight decay rate of $5\times 10^{-4}$, and a loss function given by Equation \ref{eq9}. Learning rate decays by a factor of 0.5 if a validation loss does not decrease in the past 10 epochs. Model training stops if a validation loss does not decrease in the past 30 epochs. For other hyper-parameters, we empirically set $d_e$ to 384 and $n_t$ to 4. All experiments run on a machine with 2 $\times$ V100s (32G) GPUs. 

\subsubsection{Baseline comparisons}
\label{subsubsec:baseline}
Deep Sets \citep{zaheer2017deep} and Attention-MIL \citep{ilse2018attn} are adopted as baselines since they are classical MIL approaches. And the approaches presented in Table \ref{tb1} are also used for comparisons, \textit{except} (i) BDOCOX \citep{shao2021weakly} and RankSurv \citep{di2020ranking}: source codes are unavailable; (ii) DSMIL \citep{li2021dt} and DTFD-MIL \citep{zhang2022dtfd}: inapplicable to survival prediction tasks (refer to Section \ref{subsubsec:amilmethod}). Moreover, WSISA \citep{zhu2017wsisa} and DT-MIL \citep{li2021dt} are not among our baselines, because the former is not a MIL model and the latter cannot be used for spatially-sparse patches. HIPT \citep{Chen2022scaling} is based on a pre-training strategy so it is also not used. Despite unavailable source codes, SeTranSurv \citep{huang2021integration} is also in our baselines, since it is the first Transformer-based prognosis model. An inhouse-implementation is used for it. 

\begin{table*}[tp]
\small
\centering
\caption{Ablation study on DSCA. Results are assessed by 5-fold cross-validation. Res. refers to Resolution.}
\label{tb4}
\begin{tabular}{cccccccc}
\toprule
\multicolumn{2}{c}{Dual-stream} & \multirow{2}{*}{Cross-attention}  & \multicolumn{3}{c}{Mean C-Index} & \multirow{2}{*}{Model Size} & \multirow{2}{*}{\# MACs} \\ \cmidrule(lr){1-2} \cmidrule(lr){4-6}
Low Res. & High Res. &  & NLST & BRCA & LGG & & \\ \midrule
\ding{51} & - & -   & 0.66174 & 0.60193 & 0.57553 & 3.15M & 695.39M \\
 - & \ding{51} & -   & 0.65914\textsubscript{ - 0.00260} & 0.60820\textsubscript{ + 0.00627} & 0.65614\textsubscript{ + 0.08061} & 1.58M & 1.61G \\
\ding{51} & \ding{51} & - & 0.66887\textsubscript{ + 0.00973} & 0.59599\textsubscript{ - 0.01221} & 0.68965\textsubscript{ + 0.03351} & 5.32M & 2.43G \\
\ding{51} & \ding{51} & \ding{51} & 0.66803\textsubscript{ - 0.00084} & 0.61224\textsubscript{ + 0.01625} & 0.70181\textsubscript{ + 0.01216} & 6.31M & 2.51G                  \\
\bottomrule
\end{tabular}
\end{table*}

\begin{table*}[tp]
\small
\centering
\caption{Comparison of the different token embedding layers in the high-resolution stream.}
\label{tb5}
\begin{tabular}{cccccc}
\toprule
\multirow{2}{*}{$\rho$}    & \multicolumn{3}{c}{Mean C-Index} & \multirow{2}{*}{Model Size} & \multirow{2}{*}{\# MACs} \\ \cmidrule(lr){2-4}
  & NLST & BRCA & LGG & & \\ \midrule
MLP & 0.66803 & 0.61224 & 0.70181 & 6.31M & 2.51G \\
Conv2D & 0.66981\textsubscript{ + 0.00178} & 0.58552\textsubscript{ - 0.02672} & 0.67344\textsubscript{ - 0.02837} & 9.45M & 13.08G \\
\bottomrule
\end{tabular}
\end{table*}

\subsection{Prognosis prediction}
\subsubsection{Overall performance}
The results of model performance are shown in Table \ref{tb3}. From this table, we can summarize that (i) our DSCA scheme achieves the best performances in C-Index on three datasets, and it shows obvious advantages on NLST and BRCA; (ii) our DSCA model generally has more learnable parameters but fewer computational costs. Furthermore, from Table \ref{tb3} we have the following empirical analysis. 

(1) We can notice that the classification-oriented model, TransMIL, is competitive with existing models. Because TransMIL specially devises a convolution-based positional encoding module to extract discriminative global features, as described in \citep{shao2021transmil}. However, this module usually brings huge computational costs (9.55G MACs, see Table \ref{tb3}). By contrast, our DSCA has an obviously-lower computational cost (2.51G MACs), owing to the square pooling layer which largely decreases the number of high-resolution patches while exploiting fine-grained features. Moreover, our scheme achieves better performance than TransMIL, especially on NLST and BRCA. In addition, it is observed that our DSCA network has more model parameters (6.31M) than TransMIL (2.67M), which indicates that our model has a higher space complexity (or memory consumption). This is largely due to the low-resolution stream of DSCA. Please refer to our ablation study for more details. 

(2) As a multi-resolution scheme similar to our DSCA, the classification-oriented H$^2$MIL performs not well in cancer prognosis. A possible reason is that H$^2$MIL has too few learnable model parameters (860.18K) to handle a more complex task, cancer prognosis. Besides that, we can notice that, although H$^2$MIL has few model parameters, its dense cross-resolution graph edges (see Figure \ref{fig0b}(b)) incur a high computation load (4.94G). Instead of explicitly using cross-resolution connections like H$^2$MIL, our DSCA adopts cross-attention to deal with the potential semantical gap between dual-resolution patches. We will dive into the core of DSCA through ablation studies. 

(3) Our DSCA gains a C-Index improvement of 6.7\%, 5.9\%, and 1.1\% on NLST, BRCA, and LGG, respectively, in comparison with SOTA baselines. This empirical result demonstrates that the proposed DSCA could often outperform existing mainstream approaches in WSI cancer prognosis. 

\subsubsection{Risk stratification}
\begin{figure}[htp]
\centering
\includegraphics[width=0.48\textwidth]{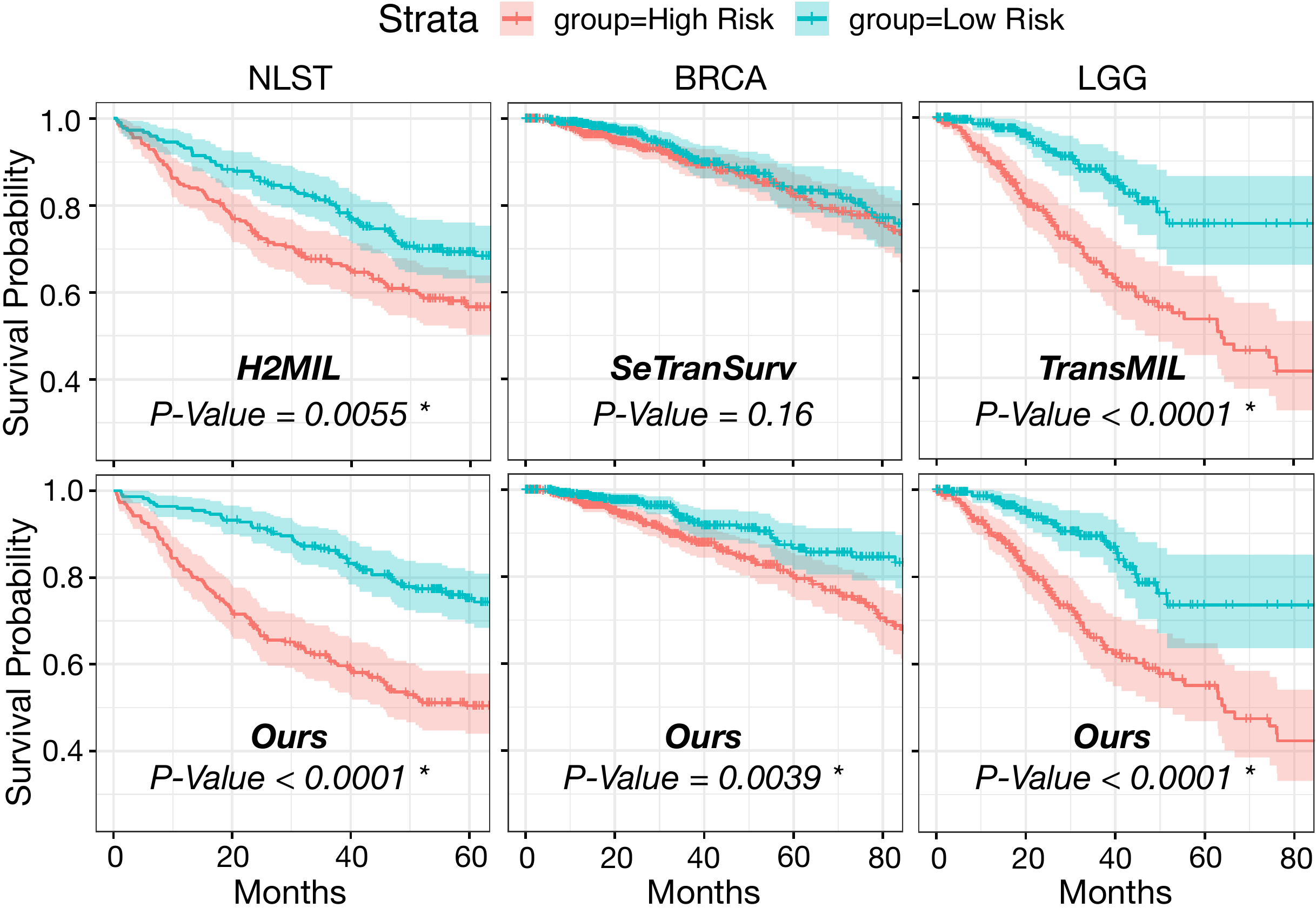}
\caption{Risk stratification. Our method and the other best ones (in C-Index) are shown. The Kaplan-Meier survival curves are plotted using ground-truth labels. The P-Value is computed by a logrank test to measure the significance of survival differences between two risk groups.}
\label{fig3}
\end{figure}

We also assess the model's ability in risk stratification, which is also a widely-used evaluation for survival analysis models. Specifically, we first aggregate out-of-sample risk predictions from the validation folds of a single dataset, and then stratify patients into two risk groups. Patients belong to a high-risk group if their risk predictions are higher than the median; otherwise, a low-risk one. Finally, we utilize the ground truth of patient survival time to plot the survival distributions of two risk groups, as shown in the Kaplan-Meier survival curves of Figure \ref{fig3}, and use a logrank test to measure if the difference of two survival distributions is statistically significant (P-Value $<$ 0.05). A survival model is of good risk discrimination if there is a significant difference between high- and low-risk groups.

From Figure \ref{fig3}, we can see that our models, trained on three cancer types separately, can always classify patients into a high- or low-risk group with significant differences (P-Value$<$0.01). It means that our models could accurately identify poor or good prognoses from the WSIs of cancer patients. Moreover, the stratification differences given by our method are clearer than that given by others on NLST and BRCA, seen from P-Values.

\subsection{Ablation study} 

\subsubsection{Dual-stream and cross-attention}
To study how the key components, dual-stream and cross-attention, affect DSCA's performance, we start from a single-stream version that only processes single-resolution patches (\textit{i.e.}, in which one sub-stream is removed from DSCA). Note that, for the DSCA only with a high-resolution stream, we use a simple function, mean-pooling, to implement its square pooling layer. Then, we improve single-stream networks by completing dual-stream and further replacing mean-pooling with our cross-attention pooling. The results are shown in Table \ref{tb4}. 

From Table \ref{tb4}, we can observe the following empirical facts. (i) Comparing the high-resolution stream with a low-resolution one, the C-Index on LGG has a clear rise ($\sim$0.08), and the C-Index on NLST and BRCA have no obvious changes (the highest is $\sim$0.006). (ii) Transforming high-resolution stream into dual-stream, the C-Index on both LGG and NLST have improvements, and the improvement on LGG is more apparent ($\sim$0.03); whereas the C-Index on BRCA drops by $\sim$0.01 (this drop is fixed afterward by adding cross-attention). (iii) After applying cross-attention pooling, C-Index performance can be further boosted, except for a slight drop ($<$0.001) on NLST. (iv) Moreover, from Table \ref{tb4}, we can see that the model size of DSCA is greatly increased (from 1.58M to 6.31M) compared to the beginning version. This is largely due to the involvement of Conv1D. Despite the increase of model size in DSCA, we notice that the amount of MACs remains relatively small (1.61G--2.51G), owing to our square pooling layer, especially for enormous high-resolution patches. 

Finally, by combining these two vital components, our DSCA could often achieve the best performances on three datasets. This ablation study suggests that our key idea, dual-stream with cross-attention, indeed contributes to DSCA. 

\begin{figure*}[htp]
\centering
\includegraphics[width=0.98\textwidth]{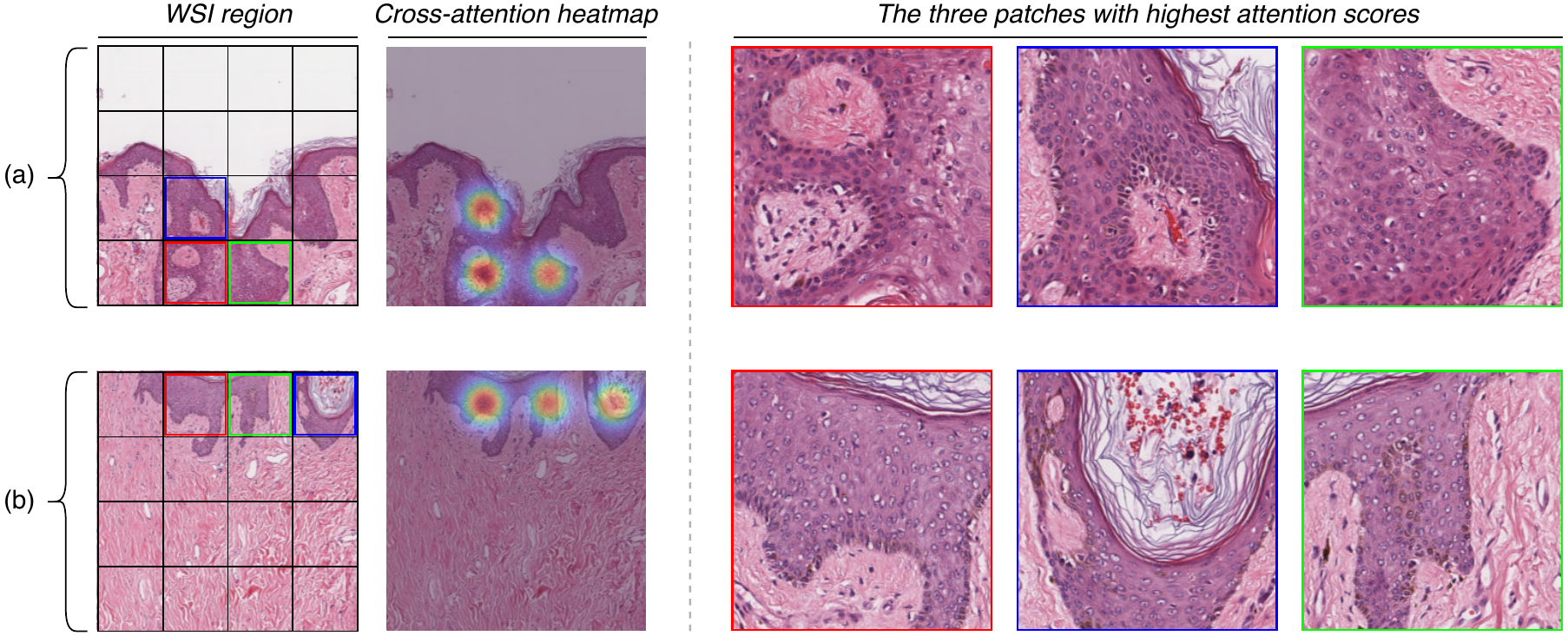}
\caption{Visualization of cross-attention maps on the region (a) and (b). Patches are exhibited in descending order according to their attention scores. Each patch region (left) is at 5$\times$ magnification and with 256 $\times$ 256 pixels. All the three selected patches are at 20$\times$ magnification and with the same scale.}
\label{fig4}
\end{figure*}
\begin{figure*}[htp]
\centering
\includegraphics[width=0.90\textwidth]{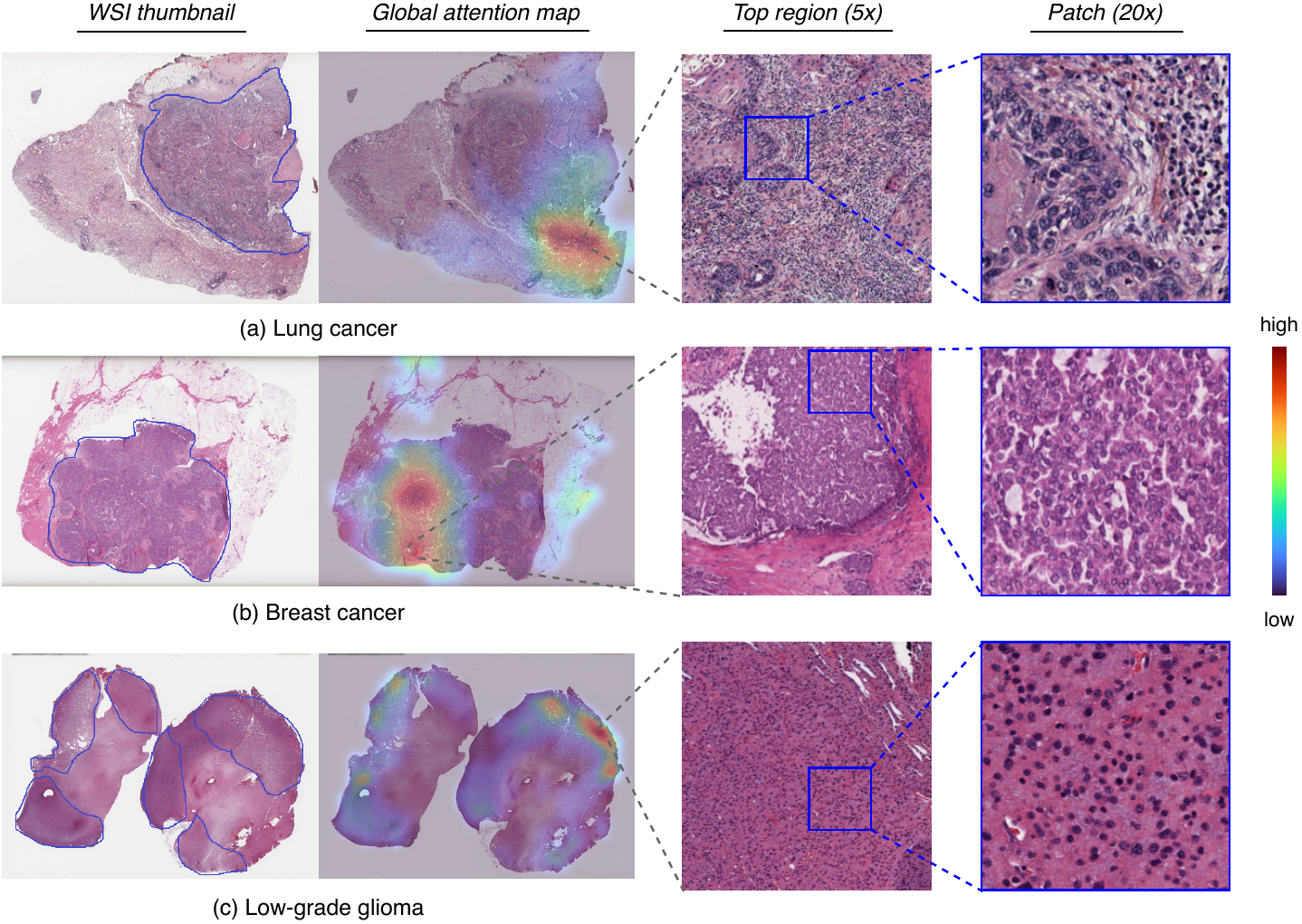}
\caption{Visualization of global attention maps on the three WSIs with different cancer diseases. The blue lines in WSI thumbnails are the annotations of the tumorous region of interest. The 50\% of regions with lower attention are assigned with zero scores for a better view. Top region means the region with high attention.}
\label{fig5}
\end{figure*}

\subsubsection{High-resolution token embedding}
We compare the two different implementations of the high-resolution token embedding layer ($\rho$), \textit{i.e.}, MLP and Conv2D, as discussed in Section \ref{subsubsec:hstream}. Comparison results are shown in Table \ref{tb5}. From these results, we can summarize that the high-resolution token embedding implemented by Conv2D cannot always bring apparent positive effects on model performance, as it only has an ignorable improvement ($<$0.002) on NLST; instead, it largely increases the model size by 3.14M and the \# MACs by 10.57G, in comparison with MLP. 

These facts indicate that the way of overlapping embedding in Conv2D may not be suitable for high-resolution WSI patches though it has been demonstrated to be effective for natural images \citep{wang2021pyramid}. By contrast, the way of compact token embedding in MLP is almost always shown to be better in both model performance and computation efficiency, which exactly demonstrates our argument (see Section \ref{subsubsec:hstream}). 

\subsubsection{Network hyper-parameters}
We test the effect of important network hyper-parameters on model performance and show its results as follows. 

(1) Sparse positional embedding (PE). The details of sparse PE can be found in Section \ref{subsubsec:pe}. We examine the DSCA network with or without sparse PE, and show their results in Table \ref{tb6}. From this table, we can find that the network with sparse PE achieves better performances on both NLST and BRCA; on LGG, the network without sparse PE is better. 

\begin{table}[htbp]
\small
\centering
\caption{Effect of sparse positional embedding (PE) on DSCA.}
\label{tb6}
\begin{tabular}{cccc}
\toprule
\multirow{2}{*}{Sparse PE} & \multicolumn{3}{c}{Mean C-Index} \\ \cmidrule(lr){2-4}
  & NLST   & BRCA   & LGG  \\ \midrule
\ding{51} & 0.66803 & 0.61224 & 0.68740  \\
\ding{55} & 0.65901 & 0.59148 & 0.70181 \\ \bottomrule
\end{tabular}
\end{table}

(2) Stream feature fusion (FF). We also examine different implementations for the feature fusion of dual-stream tokens, concatenating and adding. From the results shown in Table \ref{tb7}, we see that only on NLST, the way of adding leads that of concatenating by a quite narrow margin ($\sim$0.004). Thus we set the way of stream FF to concatenating for all experiments. 

\begin{table}[htbp]
\small
\centering
\caption{Effect of different feature fusion (FF) implementations on DSCA.}
\label{tb7}
\begin{tabular}{cccc}
\toprule
\multirow{2}{*}{Stream FF} & \multicolumn{3}{c}{Mean C-Index} \\ \cmidrule(lr){2-4}
  & NLST   & BRCA   & LGG  \\ \midrule
concatenating  & 0.66803 & 0.61224 & 0.70181  \\
adding  & 0.67216 & 0.61115 & 0.68121 \\ \bottomrule
\end{tabular}
\end{table}

\subsection{Interpretability and visualization} 
To intuitively understand the effect of cross-attention pooling, we randomly select two regions from breast cancer tissues and visualize their cross-attention results. Specifically, at first, we obtain cross-attention scores (averaged over multiple attention heads) from each 5$\times$ region and its respective 20$\times$ patches (presented as a 4$\times$4 square grid), by Equation \ref{eq5}. Then, we select the three patches with the highest attention scores, and show their fine-grained cellular details. As shown in Figure \ref{fig4}, we can observe that the cross-attention guided by global views could enable the model to focus on informative patches, \textit{e.g.}, the patches with normal or tumorous cells. Moreover, almost all the top attention patches contain many irregular cells. These cells are highly-different in size and shape (see Figure \ref{fig4}), often indicating a poor cancer prognosis \citep{alizadeh2015toward}. These visualization results show that, discriminative fine-grained features could be distilled from high-resolution patches via cross-attention pooling, thus facilitating a more accurate prognosis prediction. 

In addition, we further show the result of global attention pooling (see Section \ref{subsubsec:outlayer}, output layer), to observe which region has a high correlation with cancer prognosis. As shown in Figure \ref{fig5}, our model could capture local regions for prognosis prediction, and those predicted high-correlation regions tend to appear in RoI (region of interest) annotations. Moreover, we could see some highly-differentiated and irregular cells in these regions (5$\times$) and their enlarged patch images (20$\times$). Especially for the sample with lung cancer, there are clear differences in cell morphology between the cells located on the left side and the right side, as shown in the leftmost 20$\times$ patch in Figure \ref{fig5}(a). These observations suggest that our model could capture discriminative regions and enjoy good interpretability for cancer prognosis. 

\section{Conclusion}
This paper proposed a dual-stream network with cross-attention (DSCA) to fully utilize WSI pyramids for cancer prognosis. Unlike existing multi-resolution schemes for WSIs, we devised two sub-streams to process the patches with two resolutions. Moreover, we proposed a square pooling layer implemented by cross-attention to efficiently learn fine-grained features and make the fusion of dual-stream features smoother. A total of 3,101 WSIs from 1,911 patients is used to validate our scheme and compare our scheme with state-of-the-art approaches. Our experimental results show that: i) the key idea of dual-stream and cross-attention indeed contributes to DSCA's performance, ii) our DSCA network is more efficient than a typical multi-resolution network (H$^2$MIL), and iii) the proposed DSCA could outperform existing state-of-the-art methods in cancer prognosis by an average C-Index improvement of around 4.6\%. We hope that the proposed DSCA could encourage more works to explore WSI pyramids for broad downstream tasks, and our dual-stream architecture could serve as an effective tool for WSI-based cancer prognosis. 

In the future, we will explore other efficient strategies of multi-scale feature fusion for gigapixel WSIs, such as a reverse or bi-directional fusion of multi-resolution features. Moreover, we will study how to exploit feature pyramids directly and efficiently in WSIs, similar to the way of modern ConvNets for natural images. In the light of the development of computer vision, we believe that WSI pyramid hierarchy would become a promising research topic in computational pathology. 

\bibliographystyle{model5-names}
\bibliography{manuscript-refs.bib}

\begin{thebibliography}{54}
\expandafter\ifx\csname natexlab\endcsname\relax\def\natexlab#1{#1}\fi
\providecommand{\url}[1]{\texttt{#1}}
\providecommand{\href}[2]{#2}
\providecommand{\path}[1]{#1}
\providecommand{\DOIprefix}{doi:}
\providecommand{\ArXivprefix}{arXiv:}
\providecommand{\URLprefix}{URL: }
\providecommand{\Pubmedprefix}{pmid:}
\providecommand{\doi}[1]{\href{http://dx.doi.org/#1}{\path{#1}}}
\providecommand{\Pubmed}[1]{\href{pmid:#1}{\path{#1}}}
\providecommand{\bibinfo}[2]{#2}
\ifx\xfnm\relax \def\xfnm[#1]{\unskip,\space#1}\fi
%Type = Article
\bibitem[{Adelson et~al.(1984)Adelson, Anderson, Bergen, Burt \&
  Ogden}]{adelson1984pyramid}
\bibinfo{author}{Adelson, E.~H.}, \bibinfo{author}{Anderson, C.~H.},
  \bibinfo{author}{Bergen, J.~R.}, \bibinfo{author}{Burt, P.~J.}, \&
  \bibinfo{author}{Ogden, J.~M.} (\bibinfo{year}{1984}).
\newblock \bibinfo{title}{Pyramid methods in image processing}.
\newblock {\it \bibinfo{journal}{RCA engineer}\/},  {\it
  \bibinfo{volume}{29}\/}, \bibinfo{pages}{33--41}.
%Type = Article
\bibitem[{Alizadeh et~al.(2015)Alizadeh, Aranda, Bardelli, Blanpain, Bock,
  Borowski, Caldas, Califano, Doherty, Elsner et~al.}]{alizadeh2015toward}
\bibinfo{author}{Alizadeh, A.~A.}, \bibinfo{author}{Aranda, V.},
  \bibinfo{author}{Bardelli, A.}, \bibinfo{author}{Blanpain, C.},
  \bibinfo{author}{Bock, C.}, \bibinfo{author}{Borowski, C.},
  \bibinfo{author}{Caldas, C.}, \bibinfo{author}{Califano, A.},
  \bibinfo{author}{Doherty, M.}, \bibinfo{author}{Elsner, M.} et~al.
  (\bibinfo{year}{2015}).
\newblock \bibinfo{title}{Toward understanding and exploiting tumor
  heterogeneity}.
\newblock {\it \bibinfo{journal}{Nature medicine}\/},  {\it
  \bibinfo{volume}{21}\/}, \bibinfo{pages}{846--853}.
%Type = Article
\bibitem[{Ba et~al.(2016)Ba, Kiros \& Hinton}]{ba2016layer}
\bibinfo{author}{Ba, J.~L.}, \bibinfo{author}{Kiros, J.~R.}, \&
  \bibinfo{author}{Hinton, G.~E.} (\bibinfo{year}{2016}).
\newblock \bibinfo{title}{Layer normalization}.
\newblock {\it \bibinfo{journal}{arXiv preprint arXiv:1607.06450}\/}, .
%Type = Article
\bibitem[{Bankhead et~al.(2017)Bankhead, Loughrey, Fern{\'a}ndez, Dombrowski,
  McArt, Dunne, McQuaid, Gray, Murray, Coleman et~al.}]{bankhead2017qupath}
\bibinfo{author}{Bankhead, P.}, \bibinfo{author}{Loughrey, M.~B.},
  \bibinfo{author}{Fern{\'a}ndez, J.~A.}, \bibinfo{author}{Dombrowski, Y.},
  \bibinfo{author}{McArt, D.~G.}, \bibinfo{author}{Dunne, P.~D.},
  \bibinfo{author}{McQuaid, S.}, \bibinfo{author}{Gray, R.~T.},
  \bibinfo{author}{Murray, L.~J.}, \bibinfo{author}{Coleman, H.~G.} et~al.
  (\bibinfo{year}{2017}).
\newblock \bibinfo{title}{Qupath: Open source software for digital pathology
  image analysis}.
\newblock {\it \bibinfo{journal}{Scientific reports}\/},  {\it
  \bibinfo{volume}{7}\/}, \bibinfo{pages}{1--7}.
%Type = Article
\bibitem[{Bremnes et~al.(2011)Bremnes, Dønnem, Al-Saad, Al-Shibli, Andersen,
  Sirera, Camps, Marinez \& Busund}]{roy2011role}
\bibinfo{author}{Bremnes, R.~M.}, \bibinfo{author}{Dønnem, T.},
  \bibinfo{author}{Al-Saad, S.}, \bibinfo{author}{Al-Shibli, K.},
  \bibinfo{author}{Andersen, S.}, \bibinfo{author}{Sirera, R.},
  \bibinfo{author}{Camps, C.}, \bibinfo{author}{Marinez, I.}, \&
  \bibinfo{author}{Busund, L.-T.} (\bibinfo{year}{2011}).
\newblock \bibinfo{title}{The role of tumor stroma in cancer progression and
  prognosis: Emphasis on carcinoma-associated fibroblasts and non-small cell
  lung cancer}.
\newblock {\it \bibinfo{journal}{Journal of Thoracic Oncology}\/},  {\it
  \bibinfo{volume}{6}\/}, \bibinfo{pages}{209--217}.
  \DOIprefix\doi{https://doi.org/10.1097/JTO.0b013e3181f8a1bd}.
%Type = Article
\bibitem[{Carbonneau et~al.(2018)Carbonneau, Cheplygina, Granger \&
  Gagnon}]{carbonneau2018multiple}
\bibinfo{author}{Carbonneau, M.-A.}, \bibinfo{author}{Cheplygina, V.},
  \bibinfo{author}{Granger, E.}, \& \bibinfo{author}{Gagnon, G.}
  (\bibinfo{year}{2018}).
\newblock \bibinfo{title}{Multiple instance learning: A survey of problem
  characteristics and applications}.
\newblock {\it \bibinfo{journal}{Pattern Recognition}\/},  {\it
  \bibinfo{volume}{77}\/}, \bibinfo{pages}{329--353}.
%Type = Inproceedings
\bibitem[{Chen et~al.(2022)Chen, Chen, Li, Chen, Trister, Krishnan \&
  Mahmood}]{Chen2022scaling}
\bibinfo{author}{Chen, R.~J.}, \bibinfo{author}{Chen, C.}, \bibinfo{author}{Li,
  Y.}, \bibinfo{author}{Chen, T.~Y.}, \bibinfo{author}{Trister, A.~D.},
  \bibinfo{author}{Krishnan, R.~G.}, \& \bibinfo{author}{Mahmood, F.}
  (\bibinfo{year}{2022}).
\newblock \bibinfo{title}{Scaling vision transformers to gigapixel images via
  hierarchical self-supervised learning}.
\newblock In {\it \bibinfo{booktitle}{Proceedings of the IEEE/CVF Conference on
  Computer Vision and Pattern Recognition}\/} (pp.
  \bibinfo{pages}{16144--16155}).
%Type = Inproceedings
\bibitem[{Chen et~al.(2021)Chen, Lu, Shaban, Chen, Chen, Williamson \&
  Mahmood}]{chen2021whole}
\bibinfo{author}{Chen, R.~J.}, \bibinfo{author}{Lu, M.~Y.},
  \bibinfo{author}{Shaban, M.}, \bibinfo{author}{Chen, C.},
  \bibinfo{author}{Chen, T.~Y.}, \bibinfo{author}{Williamson, D.~F.}, \&
  \bibinfo{author}{Mahmood, F.} (\bibinfo{year}{2021}).
\newblock \bibinfo{title}{Whole slide images are 2d point clouds: Context-aware
  survival prediction using patch-based graph convolutional networks}.
\newblock In {\it \bibinfo{booktitle}{International Conference on Medical Image
  Computing and Computer-Assisted Intervention}\/} (pp.
  \bibinfo{pages}{339--349}).
\newblock \bibinfo{organization}{Springer}.
%Type = Article
\bibitem[{Cox(1975)}]{cox1975partial}
\bibinfo{author}{Cox, D.~R.} (\bibinfo{year}{1975}).
\newblock \bibinfo{title}{Partial likelihood}.
\newblock {\it \bibinfo{journal}{Biometrika}\/},  {\it \bibinfo{volume}{62}\/},
  \bibinfo{pages}{269--276}.
%Type = Article
\bibitem[{De~Boer et~al.(2009)De~Boer, Taskila, Ojaj{\"a}rvi, Van~Dijk \&
  Verbeek}]{de2009cancer}
\bibinfo{author}{De~Boer, A.~G.}, \bibinfo{author}{Taskila, T.},
  \bibinfo{author}{Ojaj{\"a}rvi, A.}, \bibinfo{author}{Van~Dijk, F.~J.}, \&
  \bibinfo{author}{Verbeek, J.~H.} (\bibinfo{year}{2009}).
\newblock \bibinfo{title}{Cancer survivors and unemployment: a meta-analysis
  and meta-regression}.
\newblock {\it \bibinfo{journal}{JAMA}\/},  {\it \bibinfo{volume}{301}\/},
  \bibinfo{pages}{753--762}.
%Type = Inproceedings
\bibitem[{Deng et~al.(2009)Deng, Dong, Socher, Li, Li \&
  Fei-Fei}]{deng2009imagenet}
\bibinfo{author}{Deng, J.}, \bibinfo{author}{Dong, W.},
  \bibinfo{author}{Socher, R.}, \bibinfo{author}{Li, L.-J.},
  \bibinfo{author}{Li, K.}, \& \bibinfo{author}{Fei-Fei, L.}
  (\bibinfo{year}{2009}).
\newblock \bibinfo{title}{Imagenet: A large-scale hierarchical image database}.
\newblock In {\it \bibinfo{booktitle}{2009 IEEE conference on computer vision
  and pattern recognition}\/} (pp. \bibinfo{pages}{248--255}).
\newblock \bibinfo{organization}{IEEE}.
%Type = Inproceedings
\bibitem[{Di et~al.(2020)Di, Li, Zhang \& Gao}]{di2020ranking}
\bibinfo{author}{Di, D.}, \bibinfo{author}{Li, S.}, \bibinfo{author}{Zhang,
  J.}, \& \bibinfo{author}{Gao, Y.} (\bibinfo{year}{2020}).
\newblock \bibinfo{title}{Ranking-based survival prediction on
  histopathological whole-slide images}.
\newblock In {\it \bibinfo{booktitle}{International Conference on Medical Image
  Computing and Computer-Assisted Intervention}\/} (pp.
  \bibinfo{pages}{428--438}).
\newblock \bibinfo{organization}{Springer}.
%Type = Article
\bibitem[{Dietterich et~al.(1997)Dietterich, Lathrop \&
  Lozano-P{\'e}rez}]{dietterich1997solving}
\bibinfo{author}{Dietterich, T.~G.}, \bibinfo{author}{Lathrop, R.~H.}, \&
  \bibinfo{author}{Lozano-P{\'e}rez, T.} (\bibinfo{year}{1997}).
\newblock \bibinfo{title}{Solving the multiple instance problem with
  axis-parallel rectangles}.
\newblock {\it \bibinfo{journal}{Artificial intelligence}\/},  {\it
  \bibinfo{volume}{89}\/}, \bibinfo{pages}{31--71}.
%Type = Article
\bibitem[{Ding et~al.(2022)Ding, Gao, Wang, Lu \& Shi}]{ding2022fractal}
\bibinfo{author}{Ding, S.}, \bibinfo{author}{Gao, Z.}, \bibinfo{author}{Wang,
  J.}, \bibinfo{author}{Lu, M.}, \& \bibinfo{author}{Shi, J.}
  (\bibinfo{year}{2022}).
\newblock \bibinfo{title}{Fractal graph convolutional network with mlp-mixer
  based multi-path feature fusion for classification of histopathological
  images}.
\newblock {\it \bibinfo{journal}{Expert Systems with Applications}\/},  (p.
  \bibinfo{pages}{118793}).
%Type = Inproceedings
\bibitem[{Dosovitskiy et~al.(2021)Dosovitskiy, Beyer, Kolesnikov, Weissenborn,
  Zhai, Unterthiner, Dehghani, Minderer, Heigold, Gelly, Uszkoreit \&
  Houlsby}]{dosovitskiy2020vit}
\bibinfo{author}{Dosovitskiy, A.}, \bibinfo{author}{Beyer, L.},
  \bibinfo{author}{Kolesnikov, A.}, \bibinfo{author}{Weissenborn, D.},
  \bibinfo{author}{Zhai, X.}, \bibinfo{author}{Unterthiner, T.},
  \bibinfo{author}{Dehghani, M.}, \bibinfo{author}{Minderer, M.},
  \bibinfo{author}{Heigold, G.}, \bibinfo{author}{Gelly, S.},
  \bibinfo{author}{Uszkoreit, J.}, \& \bibinfo{author}{Houlsby, N.}
  (\bibinfo{year}{2021}).
\newblock \bibinfo{title}{An image is worth 16x16 words: Transformers for image
  recognition at scale}.
\newblock In {\it \bibinfo{booktitle}{International Conference on Learning
  Representations}\/}.
\newblock \URLprefix \url{https://openreview.net/forum?id=YicbFdNTTy}.
%Type = Article
\bibitem[{Fu et~al.(2019)Fu, Liu, Lin, Deng, Hu \& Zheng}]{fu2019predict}
\bibinfo{author}{Fu, B.}, \bibinfo{author}{Liu, P.}, \bibinfo{author}{Lin, J.},
  \bibinfo{author}{Deng, L.}, \bibinfo{author}{Hu, K.}, \&
  \bibinfo{author}{Zheng, H.} (\bibinfo{year}{2019}).
\newblock \bibinfo{title}{Predicting invasive disease-free survival for early
  stage breast cancer patients using follow-up clinical data}.
\newblock {\it \bibinfo{journal}{IEEE Transactions on Biomedical
  Engineering}\/},  {\it \bibinfo{volume}{66}\/}, \bibinfo{pages}{2053--2064}.
  \DOIprefix\doi{10.1109/TBME.2018.2882867}.
%Type = Inproceedings
\bibitem[{He et~al.(2016)He, Zhang, Ren \& Sun}]{he2016deep}
\bibinfo{author}{He, K.}, \bibinfo{author}{Zhang, X.}, \bibinfo{author}{Ren,
  S.}, \& \bibinfo{author}{Sun, J.} (\bibinfo{year}{2016}).
\newblock \bibinfo{title}{Deep residual learning for image recognition}.
\newblock In {\it \bibinfo{booktitle}{Proceedings of the IEEE conference on
  computer vision and pattern recognition}\/} (pp. \bibinfo{pages}{770--778}).
%Type = Article
\bibitem[{Heagerty \& Zheng(2005)}]{heagerty2005survival}
\bibinfo{author}{Heagerty, P.~J.}, \& \bibinfo{author}{Zheng, Y.}
  (\bibinfo{year}{2005}).
\newblock \bibinfo{title}{Survival model predictive accuracy and roc curves}.
\newblock {\it \bibinfo{journal}{Biometrics}\/},  {\it \bibinfo{volume}{61}\/},
  \bibinfo{pages}{92--105}.
  \DOIprefix\doi{https://doi.org/10.1111/j.0006-341X.2005.030814.x}.
%Type = Article
\bibitem[{Hou et~al.(2022)Hou, Yu, Lin, Huang, Yu, Qin \& Wang}]{hou2022h2}
\bibinfo{author}{Hou, W.}, \bibinfo{author}{Yu, L.}, \bibinfo{author}{Lin, C.},
  \bibinfo{author}{Huang, H.}, \bibinfo{author}{Yu, R.}, \bibinfo{author}{Qin,
  J.}, \& \bibinfo{author}{Wang, L.} (\bibinfo{year}{2022}).
\newblock \bibinfo{title}{H2-mil: Exploring hierarchical representation with
  heterogeneous multiple instance learning for whole slide image analysis}.
\newblock {\it \bibinfo{journal}{Proceedings of the AAAI Conference on
  Artificial Intelligence}\/},  {\it \bibinfo{volume}{36}\/},
  \bibinfo{pages}{933--941}. \URLprefix
  \url{https://ojs.aaai.org/index.php/AAAI/article/view/19976}.
  \DOIprefix\doi{10.1609/aaai.v36i1.19976}.
%Type = Article
\bibitem[{Houssein et~al.(2021)Houssein, Emam, Ali \&
  Suganthan}]{houssein2021deep}
\bibinfo{author}{Houssein, E.~H.}, \bibinfo{author}{Emam, M.~M.},
  \bibinfo{author}{Ali, A.~A.}, \& \bibinfo{author}{Suganthan, P.~N.}
  (\bibinfo{year}{2021}).
\newblock \bibinfo{title}{Deep and machine learning techniques for medical
  imaging-based breast cancer: A comprehensive review}.
\newblock {\it \bibinfo{journal}{Expert Systems with Applications}\/},  {\it
  \bibinfo{volume}{167}\/}, \bibinfo{pages}{114161}.
%Type = Inproceedings
\bibitem[{Huang et~al.(2021)Huang, Chai, Wang, Wang, Yang \&
  Wu}]{huang2021integration}
\bibinfo{author}{Huang, Z.}, \bibinfo{author}{Chai, H.}, \bibinfo{author}{Wang,
  R.}, \bibinfo{author}{Wang, H.}, \bibinfo{author}{Yang, Y.}, \&
  \bibinfo{author}{Wu, H.} (\bibinfo{year}{2021}).
\newblock \bibinfo{title}{Integration of patch features through self-supervised
  learning and transformer for survival analysis on whole slide images}.
\newblock In {\it \bibinfo{booktitle}{International Conference on Medical Image
  Computing and Computer-Assisted Intervention}\/} (pp.
  \bibinfo{pages}{561--570}).
\newblock \bibinfo{organization}{Springer}.
%Type = Inproceedings
\bibitem[{Ilse et~al.(2018)Ilse, Tomczak \& Welling}]{ilse2018attn}
\bibinfo{author}{Ilse, M.}, \bibinfo{author}{Tomczak, J.}, \&
  \bibinfo{author}{Welling, M.} (\bibinfo{year}{2018}).
\newblock \bibinfo{title}{Attention-based deep multiple instance learning}.
\newblock In {\it \bibinfo{booktitle}{International conference on machine
  learning}\/} (pp. \bibinfo{pages}{2127--2136}).
\newblock \bibinfo{organization}{PMLR}.
%Type = Article
\bibitem[{Kandoth et~al.(2013)Kandoth, McLellan, Vandin, Ye, Niu, Lu, Xie,
  Zhang, McMichael, Wyczalkowski, Leiserson, Miller, Welch, Walter, Wendl, Ley,
  Wilson, Raphael \& Ding}]{Kandoth2013MutationalLA}
\bibinfo{author}{Kandoth, C.}, \bibinfo{author}{McLellan, M.~D.},
  \bibinfo{author}{Vandin, F.}, \bibinfo{author}{Ye, K.}, \bibinfo{author}{Niu,
  B.}, \bibinfo{author}{Lu, C.}, \bibinfo{author}{Xie, M.},
  \bibinfo{author}{Zhang, Q.}, \bibinfo{author}{McMichael, J.~F.},
  \bibinfo{author}{Wyczalkowski, M.~A.}, \bibinfo{author}{Leiserson, M. D.~M.},
  \bibinfo{author}{Miller, C.~A.}, \bibinfo{author}{Welch, J.~S.},
  \bibinfo{author}{Walter, M.~J.}, \bibinfo{author}{Wendl, M.~C.},
  \bibinfo{author}{Ley, T.~J.}, \bibinfo{author}{Wilson, R.~K.},
  \bibinfo{author}{Raphael, B.~J.}, \& \bibinfo{author}{Ding, L.}
  (\bibinfo{year}{2013}).
\newblock \bibinfo{title}{Mutational landscape and significance across 12 major
  cancer types}.
\newblock {\it \bibinfo{journal}{Nature}\/},  {\it \bibinfo{volume}{502}\/},
  \bibinfo{pages}{333 -- 339}.
%Type = Inproceedings
\bibitem[{Li et~al.(2021{\natexlab{a}})Li, Li \& Eliceiri}]{li2021dual}
\bibinfo{author}{Li, B.}, \bibinfo{author}{Li, Y.}, \&
  \bibinfo{author}{Eliceiri, K.~W.} (\bibinfo{year}{2021}{\natexlab{a}}).
\newblock \bibinfo{title}{Dual-stream multiple instance learning network for
  whole slide image classification with self-supervised contrastive learning}.
\newblock In {\it \bibinfo{booktitle}{Proceedings of the IEEE/CVF Conference on
  Computer Vision and Pattern Recognition}\/} (pp.
  \bibinfo{pages}{14318--14328}).
%Type = Inproceedings
\bibitem[{Li et~al.(2021{\natexlab{b}})Li, Yang, Zhao, Xing, Zhang, Gao, Huang,
  Wang \& Yao}]{li2021dt}
\bibinfo{author}{Li, H.}, \bibinfo{author}{Yang, F.}, \bibinfo{author}{Zhao,
  Y.}, \bibinfo{author}{Xing, X.}, \bibinfo{author}{Zhang, J.},
  \bibinfo{author}{Gao, M.}, \bibinfo{author}{Huang, J.},
  \bibinfo{author}{Wang, L.}, \& \bibinfo{author}{Yao, J.}
  (\bibinfo{year}{2021}{\natexlab{b}}).
\newblock \bibinfo{title}{Dt-mil: Deformable transformer for multi-instance
  learning on histopathological image}.
\newblock In {\it \bibinfo{booktitle}{International Conference on Medical Image
  Computing and Computer-Assisted Intervention}\/} (pp.
  \bibinfo{pages}{206--216}).
\newblock \bibinfo{organization}{Springer}.
%Type = Inproceedings
\bibitem[{Li et~al.(2018)Li, Yao, Zhu, Li \& Huang}]{li2018graph}
\bibinfo{author}{Li, R.}, \bibinfo{author}{Yao, J.}, \bibinfo{author}{Zhu, X.},
  \bibinfo{author}{Li, Y.}, \& \bibinfo{author}{Huang, J.}
  (\bibinfo{year}{2018}).
\newblock \bibinfo{title}{Graph cnn for survival analysis on whole slide
  pathological images}.
\newblock In {\it \bibinfo{booktitle}{International Conference on Medical Image
  Computing and Computer-Assisted Intervention}\/} (pp.
  \bibinfo{pages}{174--182}).
\newblock \bibinfo{organization}{Springer}.
%Type = Inproceedings
\bibitem[{Lin et~al.(2017)Lin, Doll{\'a}r, Girshick, He, Hariharan \&
  Belongie}]{lin2017feature}
\bibinfo{author}{Lin, T.-Y.}, \bibinfo{author}{Doll{\'a}r, P.},
  \bibinfo{author}{Girshick, R.}, \bibinfo{author}{He, K.},
  \bibinfo{author}{Hariharan, B.}, \& \bibinfo{author}{Belongie, S.}
  (\bibinfo{year}{2017}).
\newblock \bibinfo{title}{Feature pyramid networks for object detection}.
\newblock In {\it \bibinfo{booktitle}{Proceedings of the IEEE conference on
  computer vision and pattern recognition}\/} (pp.
  \bibinfo{pages}{2117--2125}).
%Type = Article
\bibitem[{Liu et~al.(2021)Liu, Fu, Yang, Deng, Zhong \& Zheng}]{liu2021opt}
\bibinfo{author}{Liu, P.}, \bibinfo{author}{Fu, B.}, \bibinfo{author}{Yang,
  S.~X.}, \bibinfo{author}{Deng, L.}, \bibinfo{author}{Zhong, X.}, \&
  \bibinfo{author}{Zheng, H.} (\bibinfo{year}{2021}).
\newblock \bibinfo{title}{Optimizing survival analysis of xgboost for ties to
  predict disease progression of breast cancer}.
\newblock {\it \bibinfo{journal}{IEEE Transactions on Biomedical
  Engineering}\/},  {\it \bibinfo{volume}{68}\/}, \bibinfo{pages}{148--160}.
  \DOIprefix\doi{10.1109/TBME.2020.2993278}.
%Type = Article
\bibitem[{Liu et~al.(2022)Liu, Su, Sun, Li \& Wei}]{liu2022eoca}
\bibinfo{author}{Liu, T.}, \bibinfo{author}{Su, R.}, \bibinfo{author}{Sun, C.},
  \bibinfo{author}{Li, X.}, \& \bibinfo{author}{Wei, L.}
  (\bibinfo{year}{2022}).
\newblock \bibinfo{title}{Eocsa: Predicting prognosis of epithelial ovarian
  cancer with whole slide histopathological images}.
\newblock {\it \bibinfo{journal}{Expert Systems with Applications}\/},  {\it
  \bibinfo{volume}{206}\/}, \bibinfo{pages}{117643}.
  \DOIprefix\doi{https://doi.org/10.1016/j.eswa.2022.117643}.
%Type = Article
\bibitem[{Lu et~al.(2021)Lu, Williamson, Chen, Chen, Barbieri \&
  Mahmood}]{lu2021data}
\bibinfo{author}{Lu, M.~Y.}, \bibinfo{author}{Williamson, D.~F.},
  \bibinfo{author}{Chen, T.~Y.}, \bibinfo{author}{Chen, R.~J.},
  \bibinfo{author}{Barbieri, M.}, \& \bibinfo{author}{Mahmood, F.}
  (\bibinfo{year}{2021}).
\newblock \bibinfo{title}{Data-efficient and weakly supervised computational
  pathology on whole-slide images}.
\newblock {\it \bibinfo{journal}{Nature biomedical engineering}\/},  {\it
  \bibinfo{volume}{5}\/}, \bibinfo{pages}{555--570}.
%Type = Article
\bibitem[{Moitra \& Mandal(2020)}]{moitra2020classification}
\bibinfo{author}{Moitra, D.}, \& \bibinfo{author}{Mandal, R.~K.}
  (\bibinfo{year}{2020}).
\newblock \bibinfo{title}{Classification of non-small cell lung cancer using
  one-dimensional convolutional neural network}.
\newblock {\it \bibinfo{journal}{Expert Systems with Applications}\/},  {\it
  \bibinfo{volume}{159}\/}, \bibinfo{pages}{113564}.
%Type = Incollection
\bibitem[{Pati et~al.(2020)Pati, Jaume, Fernandes, Foncubierta-Rodr{\'\i}guez,
  Feroce, Anniciello, Scognamiglio, Brancati, Riccio, Bonito
  et~al.}]{pati2020hact}
\bibinfo{author}{Pati, P.}, \bibinfo{author}{Jaume, G.},
  \bibinfo{author}{Fernandes, L.~A.},
  \bibinfo{author}{Foncubierta-Rodr{\'\i}guez, A.}, \bibinfo{author}{Feroce,
  F.}, \bibinfo{author}{Anniciello, A.~M.}, \bibinfo{author}{Scognamiglio, G.},
  \bibinfo{author}{Brancati, N.}, \bibinfo{author}{Riccio, D.},
  \bibinfo{author}{Bonito, M.~D.} et~al. (\bibinfo{year}{2020}).
\newblock \bibinfo{title}{Hact-net: A hierarchical cell-to-tissue graph neural
  network for histopathological image classification}.
\newblock In {\it \bibinfo{booktitle}{Uncertainty for Safe Utilization of
  Machine Learning in Medical Imaging, and Graphs in Biomedical Image
  Analysis}\/} (pp. \bibinfo{pages}{208--219}).
\newblock \bibinfo{publisher}{Springer}.
%Type = Article
\bibitem[{Pati et~al.(2022)Pati, Jaume, Foncubierta-Rodr{\'\i}guez, Feroce,
  Anniciello, Scognamiglio, Brancati, Fiche, Dubruc, Riccio
  et~al.}]{pati2022hier}
\bibinfo{author}{Pati, P.}, \bibinfo{author}{Jaume, G.},
  \bibinfo{author}{Foncubierta-Rodr{\'\i}guez, A.}, \bibinfo{author}{Feroce,
  F.}, \bibinfo{author}{Anniciello, A.~M.}, \bibinfo{author}{Scognamiglio, G.},
  \bibinfo{author}{Brancati, N.}, \bibinfo{author}{Fiche, M.},
  \bibinfo{author}{Dubruc, E.}, \bibinfo{author}{Riccio, D.} et~al.
  (\bibinfo{year}{2022}).
\newblock \bibinfo{title}{Hierarchical graph representations in digital
  pathology}.
\newblock {\it \bibinfo{journal}{Medical image analysis}\/},  {\it
  \bibinfo{volume}{75}\/}, \bibinfo{pages}{102264}.
%Type = Article
\bibitem[{Shao et~al.(2019)Shao, Han, Cheng, Cheng, Wang, Sun, Lu, Zhang, Zhang
  \& Huang}]{shao2019inter}
\bibinfo{author}{Shao, W.}, \bibinfo{author}{Han, Z.}, \bibinfo{author}{Cheng,
  J.}, \bibinfo{author}{Cheng, L.}, \bibinfo{author}{Wang, T.},
  \bibinfo{author}{Sun, L.}, \bibinfo{author}{Lu, Z.}, \bibinfo{author}{Zhang,
  J.}, \bibinfo{author}{Zhang, D.}, \& \bibinfo{author}{Huang, K.}
  (\bibinfo{year}{2019}).
\newblock \bibinfo{title}{Integrative analysis of pathological images and
  multi-dimensional genomic data for early-stage cancer prognosis}.
\newblock {\it \bibinfo{journal}{IEEE transactions on medical imaging}\/},
  {\it \bibinfo{volume}{39}\/}, \bibinfo{pages}{99--110}.
%Type = Article
\bibitem[{Shao et~al.(2021{\natexlab{a}})Shao, Wang, Huang, Han, Zhang \&
  Huang}]{shao2021weakly}
\bibinfo{author}{Shao, W.}, \bibinfo{author}{Wang, T.}, \bibinfo{author}{Huang,
  Z.}, \bibinfo{author}{Han, Z.}, \bibinfo{author}{Zhang, J.}, \&
  \bibinfo{author}{Huang, K.} (\bibinfo{year}{2021}{\natexlab{a}}).
\newblock \bibinfo{title}{Weakly supervised deep ordinal cox model for survival
  prediction from whole-slide pathological images}.
\newblock {\it \bibinfo{journal}{IEEE Transactions on Medical Imaging}\/},
  {\it \bibinfo{volume}{40}\/}, \bibinfo{pages}{3739--3747}.
%Type = Inproceedings
\bibitem[{Shao et~al.(2021{\natexlab{b}})Shao, Bian, Chen, Wang, Zhang, Ji \&
  Zhang}]{shao2021transmil}
\bibinfo{author}{Shao, Z.}, \bibinfo{author}{Bian, H.}, \bibinfo{author}{Chen,
  Y.}, \bibinfo{author}{Wang, Y.}, \bibinfo{author}{Zhang, J.},
  \bibinfo{author}{Ji, X.}, \& \bibinfo{author}{Zhang, Y.}
  (\bibinfo{year}{2021}{\natexlab{b}}).
\newblock \bibinfo{title}{Trans{MIL}: Transformer based correlated multiple
  instance learning for whole slide image classification}.
\newblock In \bibinfo{editor}{A.~Beygelzimer}, \bibinfo{editor}{Y.~Dauphin},
  \bibinfo{editor}{P.~Liang}, \& \bibinfo{editor}{J.~W. Vaughan} (Eds.), {\it
  \bibinfo{booktitle}{Advances in Neural Information Processing Systems}\/}.
\newblock \URLprefix \url{https://openreview.net/forum?id=LKUfuWxajHc}.
%Type = Article
\bibitem[{Simonyan \& Zisserman(2014)}]{simonyan2014very}
\bibinfo{author}{Simonyan, K.}, \& \bibinfo{author}{Zisserman, A.}
  (\bibinfo{year}{2014}).
\newblock \bibinfo{title}{Very deep convolutional networks for large-scale
  image recognition}.
\newblock {\it \bibinfo{journal}{arXiv preprint arXiv:1409.1556}\/}, .
%Type = Article
\bibitem[{Skrede et~al.(2020)Skrede, De~Raedt, Kleppe, Hveem, Liest{\o}l,
  Maddison, Askautrud, Pradhan, Nesheim, Albregtsen et~al.}]{skrede2020deep}
\bibinfo{author}{Skrede, O.-J.}, \bibinfo{author}{De~Raedt, S.},
  \bibinfo{author}{Kleppe, A.}, \bibinfo{author}{Hveem, T.~S.},
  \bibinfo{author}{Liest{\o}l, K.}, \bibinfo{author}{Maddison, J.},
  \bibinfo{author}{Askautrud, H.~A.}, \bibinfo{author}{Pradhan, M.},
  \bibinfo{author}{Nesheim, J.~A.}, \bibinfo{author}{Albregtsen, F.} et~al.
  (\bibinfo{year}{2020}).
\newblock \bibinfo{title}{Deep learning for prediction of colorectal cancer
  outcome: a discovery and validation study}.
\newblock {\it \bibinfo{journal}{The Lancet}\/},  {\it
  \bibinfo{volume}{395}\/}, \bibinfo{pages}{350--360}.
%Type = Inproceedings
\bibitem[{Tan et~al.(2020)Tan, Pang \& Le}]{tan2020efficientdet}
\bibinfo{author}{Tan, M.}, \bibinfo{author}{Pang, R.}, \& \bibinfo{author}{Le,
  Q.~V.} (\bibinfo{year}{2020}).
\newblock \bibinfo{title}{Efficientdet: Scalable and efficient object
  detection}.
\newblock In {\it \bibinfo{booktitle}{Proceedings of the IEEE/CVF conference on
  computer vision and pattern recognition}\/} (pp.
  \bibinfo{pages}{10781--10790}).
%Type = Article
\bibitem[{Team(2011)}]{nlst2011}
\bibinfo{author}{Team, N. L. S. T.~R.} (\bibinfo{year}{2011}).
\newblock \bibinfo{title}{The national lung screening trial: overview and study
  design}.
\newblock {\it \bibinfo{journal}{Radiology}\/},  {\it \bibinfo{volume}{258}\/},
  \bibinfo{pages}{243--53}.
  \DOIprefix\doi{https://doi.org/10.1148/radiol.10091808}.
%Type = Article
\bibitem[{Vaswani et~al.(2017)Vaswani, Shazeer, Parmar, Uszkoreit, Jones,
  Gomez, Kaiser \& Polosukhin}]{vaswani2017attn}
\bibinfo{author}{Vaswani, A.}, \bibinfo{author}{Shazeer, N.},
  \bibinfo{author}{Parmar, N.}, \bibinfo{author}{Uszkoreit, J.},
  \bibinfo{author}{Jones, L.}, \bibinfo{author}{Gomez, A.~N.},
  \bibinfo{author}{Kaiser, {\L}.}, \& \bibinfo{author}{Polosukhin, I.}
  (\bibinfo{year}{2017}).
\newblock \bibinfo{title}{Attention is all you need}.
\newblock {\it \bibinfo{journal}{Advances in neural information processing
  systems}\/},  {\it \bibinfo{volume}{30}\/}.
%Type = Inproceedings
\bibitem[{Wang et~al.(2021)Wang, Xie, Li, Fan, Song, Liang, Lu, Luo \&
  Shao}]{wang2021pyramid}
\bibinfo{author}{Wang, W.}, \bibinfo{author}{Xie, E.}, \bibinfo{author}{Li,
  X.}, \bibinfo{author}{Fan, D.-P.}, \bibinfo{author}{Song, K.},
  \bibinfo{author}{Liang, D.}, \bibinfo{author}{Lu, T.}, \bibinfo{author}{Luo,
  P.}, \& \bibinfo{author}{Shao, L.} (\bibinfo{year}{2021}).
\newblock \bibinfo{title}{Pyramid vision transformer: A versatile backbone for
  dense prediction without convolutions}.
\newblock In {\it \bibinfo{booktitle}{Proceedings of the IEEE/CVF International
  Conference on Computer Vision}\/} (pp. \bibinfo{pages}{568--578}).
%Type = Article
\bibitem[{Wang et~al.(2018)Wang, Yan, Tang, Bai \& Liu}]{wang2018revisiting}
\bibinfo{author}{Wang, X.}, \bibinfo{author}{Yan, Y.}, \bibinfo{author}{Tang,
  P.}, \bibinfo{author}{Bai, X.}, \& \bibinfo{author}{Liu, W.}
  (\bibinfo{year}{2018}).
\newblock \bibinfo{title}{Revisiting multiple instance neural networks}.
\newblock {\it \bibinfo{journal}{Pattern Recognition}\/},  {\it
  \bibinfo{volume}{74}\/}, \bibinfo{pages}{15--24}.
%Type = Article
\bibitem[{Yao et~al.(2020)Yao, Zhu, Jonnagaddala, Hawkins \&
  Huang}]{yao2020whole}
\bibinfo{author}{Yao, J.}, \bibinfo{author}{Zhu, X.},
  \bibinfo{author}{Jonnagaddala, J.}, \bibinfo{author}{Hawkins, N.}, \&
  \bibinfo{author}{Huang, J.} (\bibinfo{year}{2020}).
\newblock \bibinfo{title}{Whole slide images based cancer survival prediction
  using attention guided deep multiple instance learning networks}.
\newblock {\it \bibinfo{journal}{Medical Image Analysis}\/},  {\it
  \bibinfo{volume}{65}\/}, \bibinfo{pages}{101789}.
%Type = Article
\bibitem[{Yu et~al.(2016)Yu, Zhang, Berry, Altman, R{\'e}, Rubin \&
  Snyder}]{yu2016pred}
\bibinfo{author}{Yu, K.-H.}, \bibinfo{author}{Zhang, C.},
  \bibinfo{author}{Berry, G.~J.}, \bibinfo{author}{Altman, R.~B.},
  \bibinfo{author}{R{\'e}, C.}, \bibinfo{author}{Rubin, D.~L.}, \&
  \bibinfo{author}{Snyder, M.} (\bibinfo{year}{2016}).
\newblock \bibinfo{title}{Predicting non-small cell lung cancer prognosis by
  fully automated microscopic pathology image features}.
\newblock {\it \bibinfo{journal}{Nature communications}\/},  {\it
  \bibinfo{volume}{7}\/}, \bibinfo{pages}{1--10}.
%Type = Article
\bibitem[{Zadeh \& Schmid(2021)}]{zadeh2020bias}
\bibinfo{author}{Zadeh, S.~G.}, \& \bibinfo{author}{Schmid, M.}
  (\bibinfo{year}{2021}).
\newblock \bibinfo{title}{Bias in cross-entropy-based training of deep survival
  networks}.
\newblock {\it \bibinfo{journal}{IEEE Transactions on Pattern Analysis and
  Machine Intelligence}\/},  {\it \bibinfo{volume}{43}\/},
  \bibinfo{pages}{3126--3137}. \DOIprefix\doi{10.1109/TPAMI.2020.2979450}.
%Type = Article
\bibitem[{Zaheer et~al.(2017)Zaheer, Kottur, Ravanbakhsh, Poczos, Salakhutdinov
  \& Smola}]{zaheer2017deep}
\bibinfo{author}{Zaheer, M.}, \bibinfo{author}{Kottur, S.},
  \bibinfo{author}{Ravanbakhsh, S.}, \bibinfo{author}{Poczos, B.},
  \bibinfo{author}{Salakhutdinov, R.~R.}, \& \bibinfo{author}{Smola, A.~J.}
  (\bibinfo{year}{2017}).
\newblock \bibinfo{title}{Deep sets}.
\newblock {\it \bibinfo{journal}{Advances in neural information processing
  systems}\/},  {\it \bibinfo{volume}{30}\/}.
%Type = Article
\bibitem[{Zarella et~al.(2018)Zarella, Bowman;, Aeffner, Farahani, Xthona;,
  Absar, Parwani, Bui \& Hartman}]{zarella2018apractical}
\bibinfo{author}{Zarella, M.~D.}, \bibinfo{author}{Bowman;, D.},
  \bibinfo{author}{Aeffner, F.}, \bibinfo{author}{Farahani, N.},
  \bibinfo{author}{Xthona;, A.}, \bibinfo{author}{Absar, S.~F.},
  \bibinfo{author}{Parwani, A.}, \bibinfo{author}{Bui, M.}, \&
  \bibinfo{author}{Hartman, D.~J.} (\bibinfo{year}{2018}).
\newblock \bibinfo{title}{{A Practical Guide to Whole Slide Imaging: A White
  Paper From the Digital Pathology Association}}.
\newblock {\it \bibinfo{journal}{Archives of Pathology \& Laboratory
  Medicine}\/},  {\it \bibinfo{volume}{143}\/}, \bibinfo{pages}{222--234}.
  \DOIprefix\doi{10.5858/arpa.2018-0343-RA}.
%Type = Inproceedings
\bibitem[{Zeiler \& Fergus(2014)}]{zeiler2014visualizing}
\bibinfo{author}{Zeiler, M.~D.}, \& \bibinfo{author}{Fergus, R.}
  (\bibinfo{year}{2014}).
\newblock \bibinfo{title}{Visualizing and understanding convolutional
  networks}.
\newblock In {\it \bibinfo{booktitle}{European conference on computer
  vision}\/} (pp. \bibinfo{pages}{818--833}).
\newblock \bibinfo{organization}{Springer}.
%Type = Article
\bibitem[{Zeiser et~al.(2021)Zeiser, da~Costa, {de Oliveira Ramos}, Bohn,
  Santos \& Roehe}]{zeiser2021deepbatch}
\bibinfo{author}{Zeiser, F.~A.}, \bibinfo{author}{da~Costa, C.~A.},
  \bibinfo{author}{{de Oliveira Ramos}, G.}, \bibinfo{author}{Bohn, H.~C.},
  \bibinfo{author}{Santos, I.}, \& \bibinfo{author}{Roehe, A.~V.}
  (\bibinfo{year}{2021}).
\newblock \bibinfo{title}{Deepbatch: A hybrid deep learning model for
  interpretable diagnosis of breast cancer in whole-slide images}.
\newblock {\it \bibinfo{journal}{Expert Systems with Applications}\/},  {\it
  \bibinfo{volume}{185}\/}, \bibinfo{pages}{115586}.
  \DOIprefix\doi{https://doi.org/10.1016/j.eswa.2021.115586}.
%Type = Inproceedings
\bibitem[{Zhang et~al.(2020)Zhang, Zhang, Tang, Wang, Hua \&
  Sun}]{zhang2020feature}
\bibinfo{author}{Zhang, D.}, \bibinfo{author}{Zhang, H.},
  \bibinfo{author}{Tang, J.}, \bibinfo{author}{Wang, M.}, \bibinfo{author}{Hua,
  X.}, \& \bibinfo{author}{Sun, Q.} (\bibinfo{year}{2020}).
\newblock \bibinfo{title}{Feature pyramid transformer}.
\newblock In {\it \bibinfo{booktitle}{European Conference on Computer
  Vision}\/} (pp. \bibinfo{pages}{323--339}).
\newblock \bibinfo{organization}{Springer}.
%Type = Inproceedings
\bibitem[{Zhang et~al.(2022)Zhang, Meng, Zhao, Qiao, Yang, Coupland \&
  Zheng}]{zhang2022dtfd}
\bibinfo{author}{Zhang, H.}, \bibinfo{author}{Meng, Y.}, \bibinfo{author}{Zhao,
  Y.}, \bibinfo{author}{Qiao, Y.}, \bibinfo{author}{Yang, X.},
  \bibinfo{author}{Coupland, S.~E.}, \& \bibinfo{author}{Zheng, Y.}
  (\bibinfo{year}{2022}).
\newblock \bibinfo{title}{Dtfd-mil: Double-tier feature distillation multiple
  instance learning for histopathology whole slide image classification}.
\newblock In {\it \bibinfo{booktitle}{Proceedings of the IEEE Conference on
  Computer Vision and Pattern Recognition}\/} (pp.
  \bibinfo{pages}{18802--18812}).
%Type = Inproceedings
\bibitem[{Zhou et~al.(2016)Zhou, Khosla, Lapedriza, Oliva \&
  Torralba}]{zhou2016learning}
\bibinfo{author}{Zhou, B.}, \bibinfo{author}{Khosla, A.},
  \bibinfo{author}{Lapedriza, A.}, \bibinfo{author}{Oliva, A.}, \&
  \bibinfo{author}{Torralba, A.} (\bibinfo{year}{2016}).
\newblock \bibinfo{title}{Learning deep features for discriminative
  localization}.
\newblock In {\it \bibinfo{booktitle}{Proceedings of the IEEE conference on
  computer vision and pattern recognition}\/} (pp.
  \bibinfo{pages}{2921--2929}).
%Type = Inproceedings
\bibitem[{Zhu et~al.(2017)Zhu, Yao, Zhu \& Huang}]{zhu2017wsisa}
\bibinfo{author}{Zhu, X.}, \bibinfo{author}{Yao, J.}, \bibinfo{author}{Zhu,
  F.}, \& \bibinfo{author}{Huang, J.} (\bibinfo{year}{2017}).
\newblock \bibinfo{title}{Wsisa: Making survival prediction from whole slide
  histopathological images}.
\newblock In {\it \bibinfo{booktitle}{Proceedings of the IEEE Conference on
  Computer Vision and Pattern Recognition}\/} (pp.
  \bibinfo{pages}{7234--7242}).

\end{thebibliography}

\end{document}